\documentclass[aps,pra,twocolumn,superscriptaddress,showpacs,halfspacing,10pt]{revtex4-1}

\usepackage{longtable}
\usepackage{graphicx}
\usepackage{todonotes}
\usepackage{braket}
\usepackage{amsmath}
\usepackage{upgreek}
\usepackage{here}
\usepackage{float}

\begin{document}
\title{Optical spectroscopy of complex open 4$d$-shell ions Sn$^{7+}$-Sn$^{10+}$}
\author{F.~Torretti}
\affiliation{Advanced Research Center for Nanolithography, Science Park~110, 1098~XG Amsterdam, The Netherlands} \email{f.torretti@arcnl.nl}
\affiliation{Department of Physics and Astronomy, and LaserLaB, Vrije Universiteit, De Boelelaan 1081, 1081 HV Amsterdam, The Netherlands}
\author{A.~Windberger}
\affiliation{Advanced Research Center for Nanolithography, Science Park~110, 1098~XG Amsterdam, The Netherlands}
\affiliation{Max-Planck-Institut f\"ur Kernphysik, Saupfercheckweg 1, 69117 Heidelberg, Germany}
\author{A.~Ryabtsev}
\affiliation{Institute of Spectroscopy, Russian Academy of Sciences, Troitsk, Moscow, 108840 Russia}
\affiliation{EUV Labs, Ltd., Troitsk, Moscow, 108840 Russia}
\author{S.~Dobrodey}
\affiliation{Max-Planck-Institut f\"ur Kernphysik, Saupfercheckweg 1, 69117 Heidelberg, Germany}
\author{H.~Bekker}
\affiliation{Max-Planck-Institut f\"ur Kernphysik, Saupfercheckweg 1, 69117 Heidelberg, Germany}
\author{W.~Ubachs}
\affiliation{Advanced Research Center for Nanolithography, Science Park~110, 1098~XG Amsterdam, The Netherlands}
\affiliation{Department of Physics and Astronomy, and LaserLaB, Vrije Universiteit, De Boelelaan 1081, 1081 HV Amsterdam, The Netherlands}
\author{R.~Hoekstra}
\affiliation{Advanced Research Center for Nanolithography, Science Park~110, 1098~XG Amsterdam, The Netherlands}
\affiliation{Zernike Institute for Advanced Materials, University of Groningen, Nijenborgh 4, 9747 AG Groningen, The Netherlands}
\author{E.~V.~Kahl}
\author{J.~C.~Berengut}
\affiliation{School of Physics, University of New South Wales, Sydney 2052, Australia}
\author{J.~R.~Crespo~L\'{o}pez-Urrutia}
\affiliation{Max-Planck-Institut f\"ur Kernphysik, Saupfercheckweg 1, 69117 Heidelberg, Germany}
\author{O.~O.~Versolato}
\affiliation{Advanced Research Center for Nanolithography, Science Park~110, 1098~XG Amsterdam, The Netherlands}

\newcommand{\ambit}{\textsc{amb}{\footnotesize i}\textsc{t}}

\begin{abstract}
We analyze the complex level structure of ions with many-valence-electron open [Kr]\,4$d^\textrm{m}$ sub-shells ($\textrm{m}$=7--4) with \emph{ab initio} calculations based on configuration-interaction many-body perturbation theory (CI+MBPT). Charge-state-resolved optical and extreme ultraviolet (EUV) spectra of Sn$^{7+}$-Sn$^{10+}$ ions were obtained using an electron beam ion trap. Semi-empirical spectral fits carried out with the orthogonal parameters technique and \textsc{cowan} code calculations lead to 90 identifications of magnetic-dipole transitions and the determination of 79 energy ground-configuration levels, questioning some earlier EUV-line assignments. Our results, the most complete data set available to date for these ground configurations, confirm the \emph{ab initio} predictive power of CI+MBPT calculations for the these complex electronic systems.
\end{abstract}

\maketitle

\section{Introduction}
The electronic structure [Kr]\,4$d^\textrm{m}$ ($\textrm{m}$=7--4) of the highly charged ions (HCI) Sn$^{7+}$--Sn$^{10+}$ is extremely complicated due to the many electrons that occupy their open 4$d$ sub-shell, and remains inaccessible to even some of the most advanced atomic theories. Furthermore, the unresolved transition arrays \cite{Bauche1988transition} formed by the Sn ions are particularly useful for the production of 13.5-nm-wavelength extreme ultraviolet (EUV) radiation for nanolithographic applications \cite{Fujioka2008,Benschop2008,Mizoguchi2010,Banine2011}. Unfortunately, experimental assessments \cite{Azarov1993,tolstikhina2006ATOMICDATA,Svendsen1994,sugar1991resonance,sugar1992rb,ohashi2009complementary,DArcy2009transitions,DArcy2009identification,ohashi2010euv,yatsurugi2011euv,Windberger2016,Churilov2006SnVIII,Churilov2006SnIX--SnXII,Churilov2006SnXIII--XV,ryabtsev2008SnXIV} of spectral data are hampered by the prevalence of strong configuration interaction contributions, and by a high density of states which approaches the quantum-chaos regime for high excitation energies \cite{flambaum1994structure,dzuba2012chaos,berengut2015level,harabati2016electron}. In a recent study \cite{Windberger2016}, we found evidence calling for a revision of earlier identifications \cite{Churilov2006SnXIII--XV,ryabtsev2008SnXIV} in Sn$^{11+}$--Sn$^{14+}$ ions having 3 to 0 electrons in their 4$d$ sub-shell, and successfully demonstrated the suitability of Fock space coupled cluster (FSCC) calculations for systems with up to two valence electrons or holes. We now investigate other charge states relevant for the EUV production in plasmas, namely Sn$^{7+}$--Sn$^{10+}$. 

We focus on optical spectroscopy in the present work, which can resolve the complex manifold fine-structure splittings of these ions. Therefore, the analysis of optical transitions in heavy multi-electron, open-shell ions enables the most stringent tests of \emph{ab initio} atomic-structure calculations of strongly correlated systems with non-negligible many-electron Breit contributions. For such systems, a suitable theoretical tool is a combination of configuration interaction and many-body perturbation theory (CI+MBPT). The CI+MBPT method was first developed to very accurately treat neutral thallium as a three-valence-electron atom \cite{dzuba1996}. Since then, it has been markedly successful in treating also four- \cite{berengut06pra,berengut13prl,ong13pra,savukov_CI_2015} and even five-valence-electron \cite{berengut_five_2011} systems. However, as the number of valence electrons increases, it becomes less accurate. 
A recent extension of the CI+MBPT method, used here, includes particle-hole interaction, and improves the accuracy of the calculations \cite{berengut_CIMBPT_2016}. This makes it possible to treat systems with several vacancies which, e.g., are currently inaccessible to FSCC calculations.
\begin{figure*}[htb]
\includegraphics[width=17.8cm]{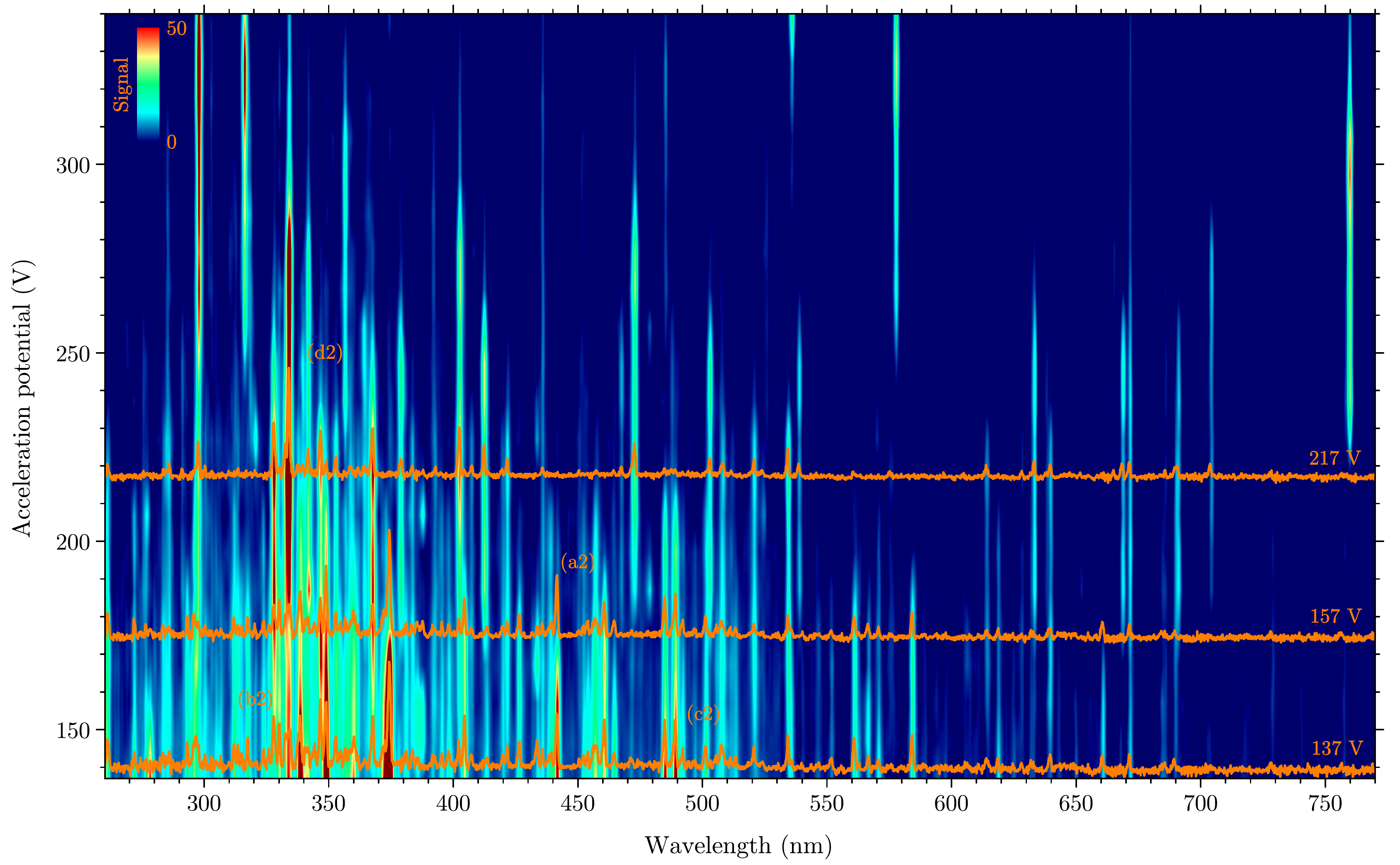}
\caption{Spectral map of Sn ions in the optical regime obtained by interpolating discrete spectra acquired at different electron beam energies (uncorrected for space charge effects). The inset color map represents the fluorescence signal strength scale in arbitrary units. The orange projections highlight spectra at three acceleration potentials at which the fluorescence of a certain charge state is highest. The lines labeled with (a2), (b2), (c2), and (d2) are shown in more detail in Fig.\,\ref{fig:chargeID}, alongside with the features recorded in the EUV (Fig.\,\ref{fig:SpectralMapEUV}) to assign the charge state.
\label{fig:SpectralMapOPT}}
\end{figure*}%
 
We present charge-state-resolved optical and EUV spectral measurements of Sn$^{7+}$--Sn$^{10+}$ ions trapped in an electron beam ion trap (EBIT), FLASH-EBIT \cite{epp2010x}, at the Max Planck Institute for Nuclear Physics (Max-Planck-Institut f\"ur Kernphysik, MPIK) in Heidelberg. EUV spectra were  obtained simultaneously with the optical ones in order to identify the charge states and assign the optical lines to them. Then, we compare the Sn$^{7+}$ data to the level structure accurately determined in Ref.\,\cite{Azarov1993}, whereby a good agreement further validates our charge state assignments. Subsequently, we perform line and level identifications for Sn$^{8+}$--Sn$^{10+}$ using semi-empirical calculations by employing the orthogonal parameters technique \cite{Uylings1993,Hansen1988a} and the \textsc{cowan} code \cite{cowan1981}. The observed Ritz combinations strongly support our semi-empirical spectral analysis. Analogous to our recent work \cite{Windberger2016}, we compare our experimental findings to previous level structure determinations from measurements of EUV spectra \cite{Churilov2006SnIX--SnXII} and find noteworthy discrepancies. Armed with these experiment-fitted level structure, we test our state-of-the-art \emph{ab initio} CI+MBPT calculations, and find them in very good agreement with the data. Both the important practical applications of the ions under study as well as the relative novelty of using CI+MBPT calculations for systems with such large numbers of valence electrons make our theory-experiment comparisons very valuable. 

\section{Experiment}
Tin ions were produced and subsequently trapped and excited using FLASH-EBIT \cite{epp2010x,Windberger2016}. In this device, the electron beam is compressed to a diameter of approximately 50\,$\upmu$m by the 6\,T magnetic field generated by a pair of superconducting coils in Helmholtz configuration. Tin atoms were brought to the trapping region by injecting a tenuous molecular beam of tetra-i-propyltin (C$_{12}$H$_{28}$Sn), which dissociated while crossing the electron beam. Tin HCI were subsequently produced through electron impact ionization, while tuning the electron beam acceleration potential allowed the selection of the desired charge states. The heavier tin HCI were trapped longitudinally by the trapping potential created using a set of drift tubes and radially by the electron beam space-charge potential, while the lighter elements in the compound (C, H) escaped from the trap. Electron collisions populate levels close to the corresponding ionization continua and profusely induce fluorescence which was recorded by two instruments: a flat-field grating spectrometer and a Czerny-Turner spectrometer for EUV and optical emissions, respectively. 

In the EUV spectrometer \cite{Bekker2015a}, light emitted by the trapped ion cloud is diffracted by a 1200\,lines/mm flat-field, grazing-incidence grating \cite{Harada1980} and recorded with a Peltier-cooled charge-coupled device (CCD) sensor. Calibration was performed using resolved bright lines of Sn in the 12--17\,nm range, for which the wavelengths were known from Ref.~\cite{Churilov2006SnIX--SnXII}, yielding a root-mean-square deviation of the calibration function residuals of 0.03\,nm. Typical observed line widths are in the order of 0.04\,nm, giving an experimental resolving power $\lambda / \delta \lambda$ of approximately 300 in the region near 13.5\,nm.

In order to measure optical spectra, FLASH-EBIT is equipped with two in-vacuo and two in-air lenses imaging the ion cloud  onto the entrance slit of a 320-mm-focal-length Czerny-Turner spectrometer equipped with a 300-lines/mm grating. For wavelength calibration Ne-Ar and Hg spectral lamps were used, depending on the spectral region. They exhibited an instrument-dominated line width of approximately 1\,nm at full-width at half-maximum (FWHM) around 400\,nm. This setup, despite its relatively low resolving power compared to typical work of the MPIK group, is very convenient for quickly covering the whole optical range in these cases where no data were available. 

A typical acquisition cycle consisted of a short calibration of the optical spectrometer, and a series of 30-minute-long simultaneous exposures of both the optical and EUV spectrometers. After each acquisition the electron beam acceleration potential was increased by 10\,V, stepping from 137\,V to 477\,V at a constant beam current of 10\,mA. This low current gives rise to a modest space-charge potential correction of approximately $25$\,V \cite{Bekker2015a,Windberger2016}. The chosen range of the acceleration potential enabled the production of charge states from Sn$^{7+}$ up to at least Sn$^{14+}$ \cite{Windberger2016}. After each energy scan the grating was rotated to measure an adjacent wavelength range while keeping a certain overlap. Next, the acceleration potential was stepped through its entire range again. This procedure was performed thrice, encompassing the full accessible wavelength range from 260 to 780\,nm. Gaussian fits were carried out to determine the centroid positions of the recorded lines. Associated error bars of approximately 0.4\,nm are dominated by the calibration uncertainty \cite{Windberger2016}. Intensities are taken from the Gaussian fits and corrected for the grating efficiency.

\section{Theory}
Two calculation methods are presented in this work. First, we present dedicated \emph{ab initio} CI+MBPT calculations performed with the \ambit\ code, and benchmark them by comparison with our experimental data. Second, in order to identify the measured transitions and the associated energy levels we utilize semi-empirical calculations using orthogonal energy scaling parameters which can be tuned to fit the spectral data. We also use the \textsc{cowan} code results on weighted transition rates $gA$ to predict line strengths and branching ratios. 

\subsection{CI+MBPT}
The detailed electronic structure of Sn$^{7+}$--Sn$^{10+}$ was calculated using the \ambit\ code which combines configuration interaction and many-body perturbation theory (CI+MBPT). Full details of this method have been presented previously~\cite{berengut06pra,berengut_five_2011,berengut_CIMBPT_2016}. Here we explain some of the physics and details relevant to the current calculations of tin ions. A more formal discussion, including mathematical details, may be found in Ref.\,\cite{berengut06pra}. Atomic units ($\hbar = m_e = e = 1$) are used in this section.

We start with a Dirac-Fock (relativistic Hartree-Fock) calculation in the $V^N$ approximation. In this approximation all $N$ electrons of the tin ion are included in the self-consistency procedure, creating a Dirac-Fock potential and electron orbitals that are optimized for the [Kr]\,$4d^\textrm{m}$ ground-state configuration.
This is particularly important for this study because between $\mathrm{m} = 4$ and 7, the $4d$ orbitals pass through the half-filled sub-shell ($4d^5$), in which the exchange contribution is maximal. We will use Sn$^{9+}$ (m = 5) as a working example in the following.

A large orbital basis is formed by diagonalizing a set of B-splines \cite{johnson88pra,dzuba1998,beloy08cpc} over the Dirac-Fock operator
\begin{equation}
\label{eq:h_DF}
\hat h_\textrm{DF} = c\, \boldsymbol{\alpha}\cdot\mathbf{p} + (\beta - 1) m_e c^2 - \frac{Z}{r} + V^N(r) .
\end{equation}
The resulting basis is then ordered by energy. The lowest few valence orbitals in each wavefunction are close to their ``spectroscopic'' counterparts, while the higher energy orbitals, so-called pseudostates, include large contributions from the continuum.

We now form a set of many-body configurations for the CI method. The CI basis includes all configurations formed by allowing single and double excitations from the $4d^5$ ground-state configuration up to $8spdf$ orbitals (i.e. including $5s$ -- $8s$, $5p$ -- $8p$, $4d$ -- $8d$, and $4f$ -- $8f$ orbitals). The configurations included in CI are defined to be within a subspace here denoted $P$; all others are within its complementary subspace $Q$. For each configuration, a complete set of projections is generated, specifying the total angular momentum and projection of each  electron in the configuration. These projections are diagonalized over the $\hat J^2$ operator to obtain configuration state functions (CSFs). The CSFs are diagonal in total angular momentum, projection, and relativistic configuration, and they form the CI basis which we denote $\left| I \right>$. All CSFs corresponding to configurations in the subspace $P$ are included in CI.

The many-electron wavefunction $\psi$ is expressed as a linear combination of CSFs from the subspace $P$,
\[
\psi = \sum_{I \in P} C_I \left| I \right> ,
\]
where the $C_I$ are obtained from the matrix eigenvalue problem.
The Hamiltonian for the CI problem is
\begin{equation}
\label{eq:H_CI}
\hat H = E_\textrm{core} + \sum_{i} \hat h_\textrm{CI} + \sum_{i < j} \frac{1}{|\mathbf{r}_i - \mathbf{r}_j|} ,
\end{equation}
where the indices $i$ and $j$ run over the valence electrons only. Note that the one-body operator $\hat h_\textrm{CI}$ is not equal to the Dirac-Fock operator: $\hat h_\textrm{CI}$ has a potential term $V^{N_\textrm{core}}$ due to the core electrons only. Therefore, the basis orbitals are not eigenvalues of the one-body CI operator, which must then be included explicitly.

Because the size of the CI matrix grows rapidly with the inclusion of additional orbitals, we must account for these configurations using many-body perturbation theory. The matrix-eigenvalue equation for the combined CI+MBPT method in second-order of perturbation theory is
\begin{equation}
\label{eq:CI+MBPT}
\sum_{J\in P} \left( H_{IJ} + \sum_{M\in Q} \frac{\left< I \right| \hat H \left| M\right>\left< M \right| \hat H \left| J\right>}{E - E_M} \right)
C_J = E C_I ,
\end{equation}
where the CSFs $\left|M\right>$ belong to configurations outside of the subspace $P$. They are, in fact, in the subspace $Q$.

Because of the extremely large number of CSFs in the subspace $Q$, it is prohibitively expensive computationally to modify all matrix elements $H_{IJ}$ directly. Instead, the CI+MBPT method includes Eq.\,\eqref{eq:CI+MBPT} by modifying the radial integrals of the one and two-body matrix elements \cite{dzuba1996}. The Slater-Condon rules for calculating matrix elements of Slater determinants ensure that this is equivalent to Eq.\,\eqref{eq:CI+MBPT}, except for the energy denominator (for a detailed discussion beyond the scope of this work, see~\cite{dzuba1996,kozlov99os,berengut06pra}). Because in this work $\hat h_\textrm{DF} \neq \hat h_\textrm{CI}$, so-called `subtraction diagrams' must be included with terms proportional to $\hat h_\textrm{CI} - \hat h_\textrm{DF}$. These diagrams can become very large when there are many valence electrons since $V^{N_\textrm{core}} - V^N $ is large, but there is cancellation between some of the largest subtraction diagrams and the three-body MBPT operator \cite{berengut_five_2011}. For this reason it is important to include three-body operators when calculating these tin ions. An alternative is to calculate the orbitals in the $V^{N-m}$ approximation (equal to $V^{N_\textrm{core}}$) as suggested in Ref.\,\cite{dzuba05pra1}; however, in this case the orbitals are much further from ``spectroscopic'', and the CI basis must be made considerably larger to correct them. In this work all one, two, and three-body second-order diagrams are included.

Until recently, only core-valence correlations were taken into account using MBPT. These correlations incorporate the effects of configurations $\left|M\right>$, which include an excitation from the $N_\textrm{core}$ electrons. It was shown in Ref.\,\cite{berengut_CIMBPT_2016} that valence-valence correlations could also be included in the same manner. Thus, in the current work, valence-valence correlations with excited orbitals up to $30spdfg$ are included; this incorporates the effect of configurations that have one or two pseudo-orbitals above $8spdf$, but have no core excitations. Furthermore, for the first time, the valence-valence subtraction diagrams presented in Ref.\,\cite{berengut_CIMBPT_2016} are also included. They vanished in that work because $\hat h_\textrm{DF}$ was the same as $\hat h_\textrm{CI}$, but play a role in the present context.

Finally, Breit and Lamb shift corrections are taken into account. The latter include the vacuum polarization (Uehling)~\cite{ginges16jpb} and self-energy~\cite{ginges16pra} corrections in the radiative potential formulation of Flambaum and Ginges~\cite{flambaum05pra}. Because both of these effects arise from the electron density near the nucleus, they have a fairly constant ratio for all the levels we calculated.

\begin{table}[H]
\caption{Energy levels of the Sn$^{9+}$ 4$d^{5}$ configuration (in cm$^{-1}$) calculated by \ambit\ CI+MBPT code. The first column give the approximate $LS$-term of the calculated energy levels. The CI values give the energy as calculated using only configuration interaction, while the $\Sigma^{\text{core}}$, $\Sigma^{\text{val}}$, Breit, and QED are the successive corrections to the CI energy by including core-valence MBPT, valence-valence MBPT, Breit, and QED contributions, respectively. The total energy including all corrections is also presented, as are the available experimentally determined values and the differences $\Delta E$ (Exp. - Total) (see main text).}
\label{tab:ambit_convergence}
\begin{ruledtabular}
\begin{tabular}{l r r r r r r r r}
&\multicolumn{8}{c}{Energy (cm$^{-1}$)}\\
\multicolumn{1}{c}{\rule[-1.75mm]{0pt}{0mm}Level}  & \multicolumn{1}{c}{CI}
&\multicolumn{1}{c}{$\Sigma^{\text{core}}$}
&\multicolumn{1}{c}{$\Sigma^{\text{val.}}$}  
&\multicolumn{1}{c}{Breit}
&\multicolumn{1}{c}{QED}  &\multicolumn{1}{c}{Total}
&\multicolumn{1}{c}{Exp} &\multicolumn{1}{c}{$\Delta E$}\\
\hline
$^6S_{5/2}$ &0	&0	&0	&0	&0	&0	&0 &0\\
$^4G_{5/2}$ &39469	&-4203 &-2141 &284  &-28  &33381	&33784 & 403\\
$^4G_{7/2}$ &42840	&-4593 &-1833 &11   &-2   &36421	&36874 &453\\
$^4G_{11/2}$ &43706	&-4676 &-1756 &-132 &3    &37145	&37535&390\\
$^4G_{9/2}$ &44212	&-4606 &-1734 &-120 &7    &37759	&38170&411\\
$^4P_{5/2}$ &43692	&-3649 &-2067 &60   &-4   &38032	&38315&283\\
$^4P_{3/2}$ &44398	&-3174 &-2316 &138  &-12  &39035	&39190&155\\
$^4P_{1/2}$ &47021	&-2711 &-2281 &32   &-1   &42060	&\\
$^4D_{7/2}$ &51789	&-4612 &-2351 &-98  &8    &44737	&44915&178\\
$^4D_{5/2}$ &55276	&-3752 &-2521 &-106 &10   &48907	&\\
$^4D_{1/2}$ &55286	&-3812 &-2310 &-190 &22   &48996	&\\
$^4D_{3/2}$ &56627	&-3340 &-2319 &-241 &25   &50753	&\\
$^2I_{11/2}$ &62330	&-7093 &-2270 &-110 &5    &52863	&53692&829\\
$^2I_{13/2}$ &65768	&-7344 &-2186 &-318 &18   &55937	&56792&855\\
$^4F_{7/2}$ &66988	&-5849 &-3102 &60   &-9   &58088	&58487&399\\
$^2D_{5/2}$ &65152	&-3732 &-2808 &-150 &18   &58479	&58756&277\\
$^4F_{3/2}$ &65795	&-4189 &-3004 &-17  &4    &58588	&58891&303\\
$^4F_{9/2}$ &67897	&-5777 &-3104 &-33  &-2   &58981	&59417&436\\
$^4F_{5/2}$ &71298	&-4933 &-2901 &-193 &20   &63291	&63643&352\\
$^2H_{9/2}$ &74999	&-5532 &-3005 &-207 &17   &66273	&66824&551\\
$^2G_{7/2}$ &75308	&-4572 &-3146 &-292 &27   &67325	&67698&373\\
$^2D_{3/2}$ &76386	&-4767 &-3007 &-325 &34   &68321	&\\
$^2F_{7/2}$ &80012	&-6786 &-3048 &-351 &32   &69859	&70199&340\\
$^2F_{5/2}$ &81165	&-5771 &-3713 &-170 &19   &71529	&71806&277\\
$^2H_{11/2}$ &82714	&-5812 &-2746 &-527 &45   &73674	&74311&637\\
$^2F_{7/2}$ &85363	&-6347 &-3651 &-323 &31   &75073	&75470&397\\
$^2G_{9/2}$ &85188	&-6283 &-3135 &-465 &42   &75347	&75795&448\\
$^2F_{5/2}$ &90363	&-7289 &-4328 &-145 &16   &78616	&78700&84\\
$^2S_{1/2}$ &87288	&-5338 &-2910 &-647 &64   &78457	&\\
$^2D_{3/2}$ &99595	&-6555 &-4503 &-149 &18   &88405	&88649&244\\
$^2D_{5/2}$ &102913	&-6472 &-4465 &-373 &37   &91640	&91927&287\\
$^2G_{9/2}$ &111086	&-8615 &-4736 &-273 &23   &97485	&98217&732\\
$^2G_{7/2}$ &112328	&-8403 &-4689 &-341 &32   &98927	&99649&722\\
\end{tabular}
\end{ruledtabular}
\end{table}

The ion Sn$^{9+}$ has a half-filled $4d$-shell, and for it the results are presented broken down into different contributions (Table\,\ref{tab:ambit_convergence}). The MBPT corrections are separated into core-valence contributions, $\Sigma^{\text{core}}$ (which correspond to unfreezing of the $4sp3d$ core), and valence-valence contributions, $\Sigma^{\text{val}}$ (introduced in Ref.\,\cite{berengut_CIMBPT_2016}), which account for configurations that include orbitals above $8spdf$. The column marked QED shows the vacuum polarization and self-energy corrections. 

The Sn$^{9+}$ and Sn$^{10+}$ ions were treated with CI+MBPT calculations using only electron excitations (the approach of Refs.\,\cite{dzuba1996,berengut06pra}). However, as the number of valence electrons increases, this electron-only approach becomes inaccurate due to very large contributions from the subtraction diagrams. To avoid this inaccuracy, the particle-hole CI+MBPT calculations are instead used for the Sn$^{7+}$ and Sn$^{8+}$ ions. This approach, described in Ref.\,\cite{berengut_CIMBPT_2016}, places the Fermi level above the $4d$ shell and treats the $4d^\mathrm{m}$ ground-state configuration as a corresponding number of valence holes in an otherwise filled shell. That is, the one-body CI operator includes the potential due to a completely filled $4d$ shell, $V^{N_\textrm{core}+10}$. Our complete CI+MBPT results for Sn$^{7+}$ and Sn$^{8+}$ in this particle-hole framework are presented in Tables\,\ref{tab:th-exp}, \ref{tab:7pambit}, \ref{tab:7plines}, \ref{tab:8p9p10p_lines}, and \ref{tab:levels} together with the results for Sn$^{9+}$ and Sn$^{10+}$.

\subsection{Orthogonal Energy Parameters} 
Line and level identifications in the Sn$^{8+}$--Sn$^{10+}$ ions were performed using the \emph{ab initio} \textsc{mcdf} (multiconfiguration Dirac-Fock) code \cite{Parpia1996}, followed by semi-empirical calculations based on the orthogonal energy scaling parameters methods for the predictions of the energy levels. The orthogonal parameters method \cite{Hansen1988a,Hansen1988b} has several advantages in comparison with the more usual Slater-Condon approximation, as for instance used in the \textsc{cowan} code \cite{cowan1981}. Firstly, the energy parameters are maximally independent, facilitating the fitting of the radial integrals of the interactions to experimental energy levels. Secondly, it is possible to include an additional number of small interactions, such as two-particle magnetic and three- and four-particle electrostatic parameters. These qualities of the orthogonal parameters method, in general, improve the agreement between calculated and measured energy levels when sufficient experimental data are available to fit its parameters. This method has been shown to be apt even for complex electronic configurations, where configuration-interaction plays a relevant role \cite{Uylings1993}. For instance, it has been applied successfully in the identification of 4$d^4$--4$d^3$5$p$ transitions in Pd$^{6+}$ \cite{Ryabtsev2012} reducing the standard deviation of the fits up to nine times compared to the \textsc{cowan} code.

Prediction of the energy levels in the 4$d^6$--4$d^4$ configuration in the Sn$^{8+}$--Sn$^{10+}$ spectra was performed by interpolation of the energy parameters between Sn$^{6+}$ (4$d^8$) \cite{VanhetHof1993, azarov1994analysis}, Sn$^{7+}$ (4$d^7$) \cite{Azarov1993}, Sn$^{11+}$ (4$d^3$) \cite{Windberger2016}, and Sn$^{12+}$ (4$d^2$) \cite{Windberger2016}. These spectra were recalculated in the framework of the orthogonal parameters to determine the scaling parameters needed for the interpolation (see also subsection \ref{ssec:parameters}). 

The transition probabilities of the magnetic dipole ($M$1) transitions were calculated using the \textsc{cowan} code. In first approximation, the \textsc{cowan} code energy levels were fitted to the energy levels predicted with the orthogonal parameters method to determine the level wavefunctions. The transition probabilities estimated with these wavefunctions were then used for the spectrum analyses and for the identification of the spectral lines. The energy levels established from the identified lines were optimized using Kramida's code \textsc{lopt} \cite{Kramida2011}. The line uncertainty relevant for the optimization was taken to be 0.4\,nm corresponding to 60\,cm$^{-1}$ at 260\,nm and 10\,cm$^{-1}$ at 600\,nm. The uncertainty was increased only for doubly classified (i.e.~lines that can be ambiguously assigned to two transitions), blended or masked lines. Final transition probabilities were obtained after the fitting of the \textsc{cowan} code to the experimentally established levels. The details of the identifications are given in the following section.

\section{Results}\label{sec:results}
\begin{figure}
\includegraphics[width=8.6cm]{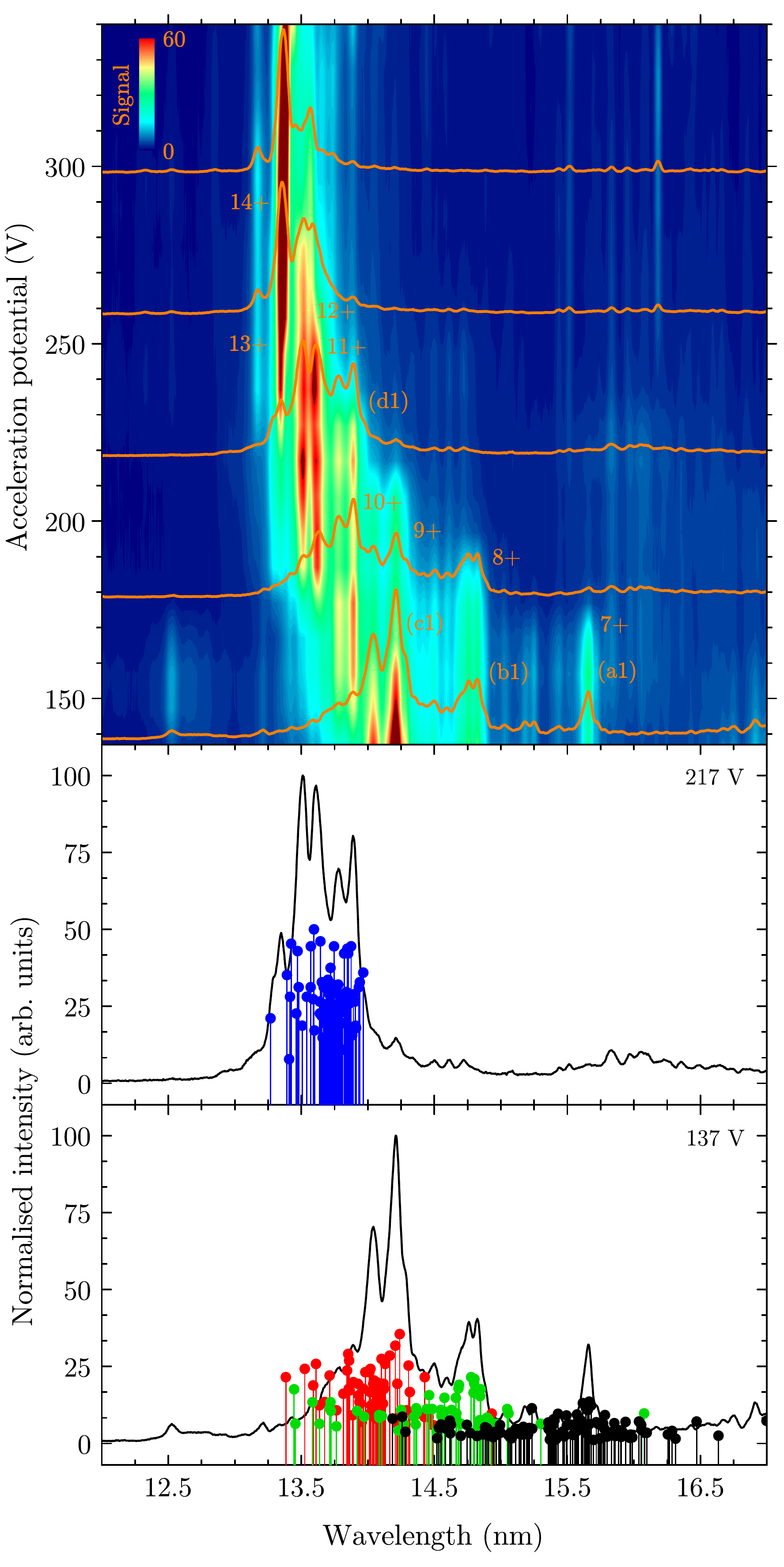}
\caption{(Upper) Spectral map of Sn ions in the extreme ultraviolet obtained by interpolating discrete spectra acquired at 10\,V acceleration potential steps starting from 137\,V (uncorrected for space charge). The inset color map indicates the fluorescence strength scale in arbitrary units. The orange individual spectra are individually scaled for visibility and spaced by approximately 40\,V, highlighting the onset of various spectral features. Labels (a1), (b1), (c1), and (d1) indicate features shown in detail in Fig.\,\ref{fig:chargeID}, where they are used for charge-state identification. (Lower) Spectra obtained at acceleration potentials 137\,V and 217\,V, individually normalized to 100 (arb. units). Scatter points represent  transitions previously observed in a vacuum spark discharge \cite{Churilov2006SnVIII,Churilov2006SnIX--SnXII}, arbitrarily scaled for visibility and to facilitate comparison (black: Sn$^{7+}$; green: Sn$^{8+}$; red: Sn$^{9+}$; blue: Sn$^{10+}$). 
\label{fig:SpectralMapEUV}}
\end{figure}

In the following, we present optical and EUV spectra of tin ions in the resolved charge states Sn$^{7+}$--Sn$^{10+}$ (see Figs.\,\ref{fig:SpectralMapOPT}, \ref{fig:SpectralMapEUV}, and \ref{fig:chargeID}). We interpret the data using orthogonal parameters and semi-empirical \textsc{cowan} code calculations, delivering the most complete data set available to date for the ground configurations of Sn$^{8+}$--Sn$^{10+}$ (with the semi-empirical results providing data also on level energies that were not directly probed experimentally). A detailed comparison of the thus obtained lines and levels with the CI+MBPT calculations is presented. Furthermore, we perform a comparison with energy levels available from existing data obtained in the EUV regime. We discuss first the charge state identification and second the line identifications. All results are summarized in Tables\,\ref{tab:th-exp}, \ref{tab:7pambit}, \ref{tab:7plines}, \ref{tab:8p9p10p_lines}, and \ref{tab:levels}. The result of line and level identifications is presented in the form of Grotrian diagrams in Fig.\,\ref{fig:Grotrian}. 

\subsection{Charge state identification}
\begin{figure*}
\includegraphics[width=17.8cm]{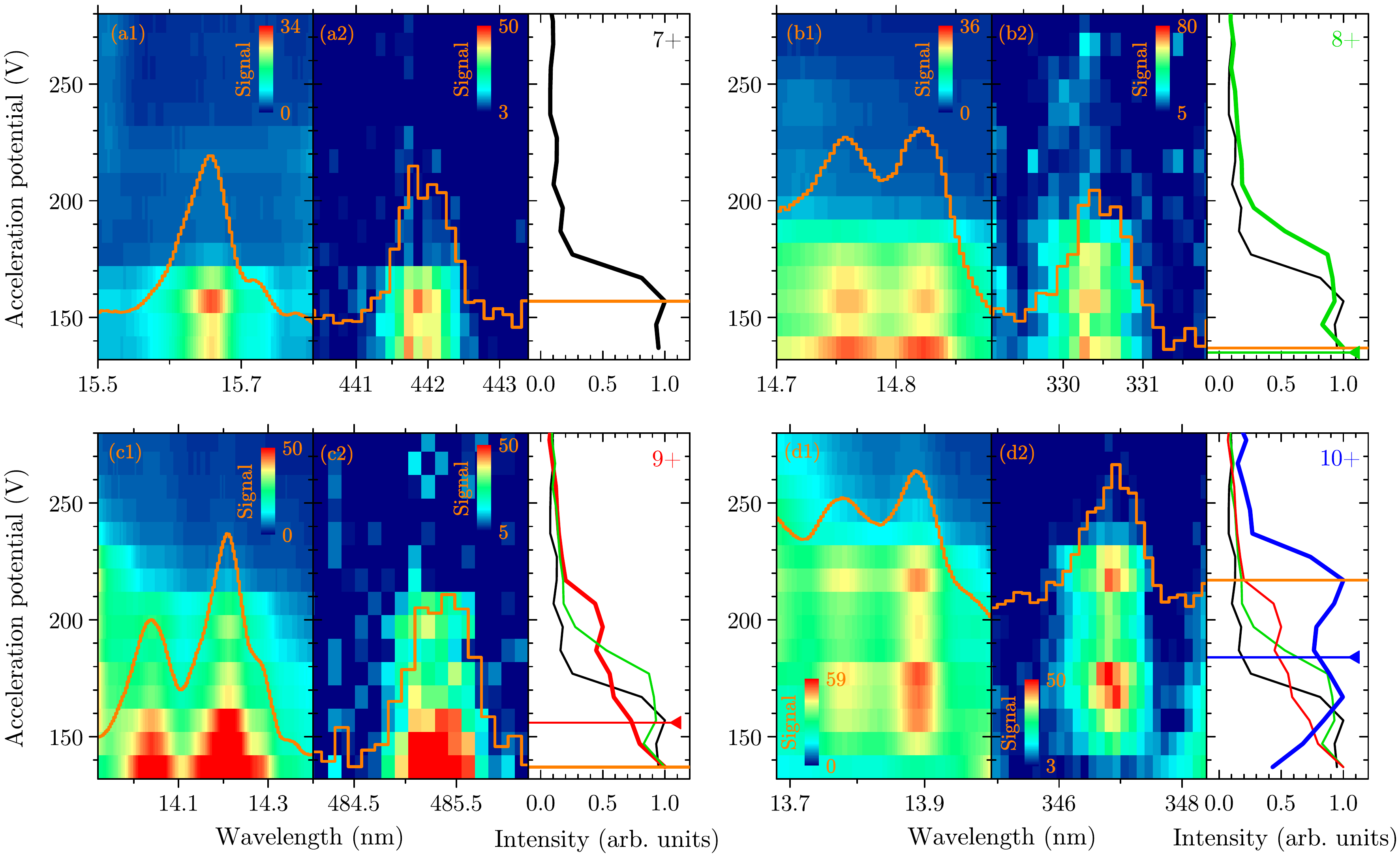}
\caption{Enlarged view of selected features from Figs.\,\ref{fig:SpectralMapOPT} and \ref{fig:SpectralMapEUV}. Independent color map scales for the fluorescence signal strength are given in arbitrary units. The fluorescence curves (black, green, red and blue lines) are determined by the averaged projections of all line intensities onto the acceleration potential axis of all the lines belonging to the same charge state, normalized to 1 at their respective maximum. Arrows indicate theoretical ionization energies (Sn$^{7+}$: 113~eV, Sn$^{8+}$: 135~eV, Sn$^{9+}$: 156~eV, and Sn$^{10+}$: 184~eV \cite{Rodrigues2004systematic,NIST_ASD}). Vertical axes show acceleration potentials (not corrected for the space-charge contribution).
\label{fig:chargeID}}
\end{figure*}

In an EBIT, charge state identification can be performed by evaluating the intensities of groups of spectral lines belonging to the same charge state as a function of the electron beam acceleration potential \cite{lopez2002visible,Bekker2015a,Windberger2016}. The doubly-peaked structure of the Sn$^{10+}$ fluorescence curve (\emph{cf.} Fig.\,\ref{fig:chargeID}), and its premature onset, has been previously observed in the optical domain for the charge states Sn$^{11+}$--Sn$^{14+}$ \cite{Windberger2016}. This phenomenon was interpreted as being caused by the existence of strongly populated \mbox{high-$J$} metastable states. They act as stepping stones for reaching the next charge state at an energy below the corresponding ionization threshold, which is derived from the ground state binding energy. Moreover, as in Ref.\,\cite{Windberger2016}, we observe that the onset of charge breeding of Sn$^{8+}$ and Sn$^{9+}$ takes place well before the respective ionization potentials are reached. Once again, this is a signature of the presence of metastable states.

In this work, the charge state identification was somewhat hampered due to the low-energy onset of the charge states Sn$^{7+}$, Sn$^{8+}$, and Sn$^{9+}$, which could not be clearly discerned in the optical data. Therefore, we relied on simultaneously obtained charge-state-resolved EUV spectra to assign optical spectra to their respective charge states. The EUV spectra were compared to previously observed clusters of lines \cite{Churilov2006SnVIII, Churilov2006SnIX--SnXII}, as shown in Fig.\,\ref{fig:SpectralMapEUV}. These lines stem from transitions to the ground configurations [Kr]\,4$d^\textrm{m}$ ($\textrm{m}$=7--4) from the 4$p^6$4$d^{\textrm{m}-1}$4$f$+4$p^5$4$d^{\textrm{m}+1}$ excited electronic configurations. Four main features have been identified in the EUV spectra that could reliably be attributed to the charge states of interest. Therefore, tracking these features as a function of the electron beam acceleration potential and comparing them to their counterparts in the optical enabled the charge-state assignment, as shown in Fig.\,\ref{fig:chargeID}. 

\subsection{Line and energy level identification}
The wavelengths and intensities of the spectral lines for each identified charge state were extracted at the acceleration potential that maximized its yield. Listed in Tables\,\ref{tab:7plines} and \ref{tab:8p9p10p_lines} are their centers and integrated intensities obtained by Gaussian fits. A direct comparison to the CI+MBPT calculations is also displayed. Deviations from the experimental data are quantified by the mean difference and standard deviation between theory and experiment for all measured transitions (see Table\,\ref{tab:th-exp}). We find very good agreement with experiment for all Sn ions studied. In the following, the results per charge state are discussed in detail.

\begin{table}[b]
\caption{Mean differences and standard deviation between our CI+MBPT calculations and experiment for measured transitions in different Sn ions (all this work).}
\label{tab:th-exp}
\begin{ruledtabular}
\begin{tabular}{l c r c}
\rule[-1.75mm]{0pt}{0mm}Ion & Configuration & \# of lines & $\Delta E_{\text{th-exp}}$ (eV) \\
\hline
Sn$^{7+}$ & 4$d^7$ & 8 & $-0.004 \pm 0.013$ \\
Sn$^{8+}$ & 4$d^6$ & 24 & $-0.005 \pm 0.023$ \\
Sn$^{9+}$ & 4$d^5$ & 30 & $-0.010 \pm 0.026$ \\
Sn$^{10+}$& 4$d^4$ & 28 & $-0.008 \pm 0.034$ \\
\end{tabular}
\end{ruledtabular}
\end{table}

\subsubsection*{Spectrum of the Sn$^{7+}$ ion}
All levels of the 4$d^7$ configuration in Sn$^{7+}$ are known from the analysis of the 4$d^7$--4$d^6$5$p$ transitions in the EUV region \cite{Azarov1993}, with estimated uncertainties of less then 12\,cm$^{-1}$. The position of the $M$1 optical transitions can be accurately obtained from the energy differences of these levels. Weighted transition rates $gA$ for these $M$1 transitions were calculated by the \textsc{cowan} code to facilitate the comparison, shown in Table\,\ref{tab:7plines}, of the eight lines measured in this work with the energy levels in Ref.\,\cite{Azarov1993}. Most of the transitions that are predicted from the available structure \cite{Azarov1993} have a relatively small calculated $gA$ value, and as such are not observable in our experiments. Three of the predicted stronger transitions (here taking $gA>35$\,s$^{-1}$), at 372.1\,nm ($^2F_{5/2}$-$^2D_{3/2}$), nearby 372.6\,nm ($^2G_{9/2}$-$^2F_{7/2}$), and at 488.4\,nm wavelength ($^2G_{7/2}$-$^2F_{7/2}$), were not reliably identified. In all three instances, this can be explained by line blending and by masking of such transitions by stronger emissions of the other charge states in the trap. The differences between our experimental wavelengths and wavelengths predicted from  Ref.\,\cite{Azarov1993} are well within mutual uncertainties, which are dominated by the 0.4\,nm uncertainty in our spectrometer calibration. Branching ratios cannot straightforwardly be used for comparison purposes, as the relevant observed transition sets (between levels 1-8/2-8, and 15-18/16-18, see Table\,\ref{tab:7plines} and Fig.\,\ref{fig:Grotrian}) are affected by blends with neighboring lines. We do not experimentally re-investigate the 4$d^7$ configuration in Sn$^{7+}$ level structure, because of the limited number of lines here well resolved and the high-accuracy and detailed results available from Ref.\,\cite{Azarov1993}. The good agreement between the present data and previous experimental observations serves as further validation of our charge state identifications.

\subsubsection*{Spectrum of the Sn$^{8+}$ ion}
The list of the identified transitions between levels within the 4$d^6$ configuration is presented in Table\,\ref{tab:8p9p10p_lines}. In total, 22 spectral lines were uniquely identified. Of these, we found 17 levels connected to the ground $^5D_4$ level. Identification of nine levels is supported by observation of 11 Ritz combinations. The level energies, optimized by \textsc{lopt} \cite{Kramida2011}, are presented in Table\,\ref{tab:levels} with their respective uncertainties. The fitting of the orthogonal parameters was performed with these optimized levels. The resulting optimized sets of parameters are given in Table\,\ref{tab:parameters}. The 354.8, 360.2, and 381.2\,nm lines are isolated lines, i.e.\ the upper and lower levels of these lines are not involved in any other transition. Therefore, the lower levels of these isolated transitions were placed at the position as calculated with the orthogonal parameters method, with an estimated uncertainty of 16\,cm$^{-1}$ (one-standard-deviation value of the orthogonal parameters fit to the experimental values). These levels were not used in the parameter fitting procedure. 

Most levels can be uniquely designated by the largest contributor in the $LS$-coupling decomposition of their wavefunctions. For example, the level labeled $^5D_3$ in Table\,\ref{tab:levels} is composed of 97\% $^5D_3$,  2\% $^3F_3$(2), and  1\% $^3D_3$. Here, the number in brackets serves to distinguish between different $LSJ$-wavefunctions designated by the same $LSJ$ values, supplementing a sequential index as defined by Nielson and Koster \cite{nielson1963}. Two exceptions are the $^1S_{0}(4)$ and $^3F_{4}(2)$ levels, which we uniquely designate by the second-largest component of the wavefunction decomposition. 

There are seven branched upper levels (numbered 6, 9, 11, 20, 21, 27, and 30 in Table\,\ref{tab:levels}, \emph{cf.} Fig.\,\ref{fig:Grotrian}). Most of the associated \textsc{cowan}-calculated branching ratios are in reasonable agreement with the experimental data, except for lines affected by blends (such as in the ratios 9-21/11-21 and 0-9/1-9) or for short-wavelength transitions below 300\,nm (featuring in the ratio 0-11/1-11 at 293.3\,nm), the intensity of which are affected by a significant drop of detection efficiency. This reduction could not be assessed in our experimental setup.

The levels of Sn$^{8+}$ found from EUV measurements on vacuum sparks \cite{Churilov2006SnIX--SnXII} belong to four disjointed groups, where levels within a single group are connected to each other by measured lines but no transitions were identified connecting the different groups. The uncertainty of the level energies within each of the groups was estimated at 10\,cm$^{-1}$, but between the groups as several 100\,cm$^{-1}$. For this reason in Table\,\ref{tab:levels}, which contains comparison of the energies of these levels ($E_{\mathrm{vs}}$) with our results ($E_\mathrm{exp}$), the four groups are identified by their respective uncertainties $z_i$ ($i$=1--4, as given in Ref.\,\cite{Churilov2006SnIX--SnXII}). These values should be interpreted as systematic common shifts, with respect to the ground state. Statistics of the differences $\Delta E_\mathrm{vs}$ (see Table\,\ref{tab:levels}) between previous identifications and the current experimental results provide a meaningful comparison between the two data sets. It is found that the two levels with shift $z_1$ are consistent with the present experimental values, with differences $\Delta E_\mathrm{vs}$ equaling $-36$ and $-22$\,cm$^{-1}$, the scatter in which is well within the uncertainty of 16\,cm$^{-1}$ obtained from the orthogonal parameter fitting of the experimental data. In contrast, the average value of $z_2$ is found to be $270 \pm 377$\,cm$^{-1}$, where the latter number represents the one-standard-deviation spread in the former. This large spread, compared to the experimental uncertainties (see Table\,\ref{tab:levels}), indicates that the respective levels from the previous work \cite{Churilov2006SnIX--SnXII} are not consistent with the present experimental values and that the classification of EUV transitions therein requires a revision. The consistency of the remaining shifts for Sn$^{8+}$, $z_3$ and $z_4$, cannot be ascertained from our data. 
	
\subsubsection*{Spectrum of the Sn$^{9+}$ ion}
\begin{figure*}[htb]
\includegraphics[width=17.8cm]{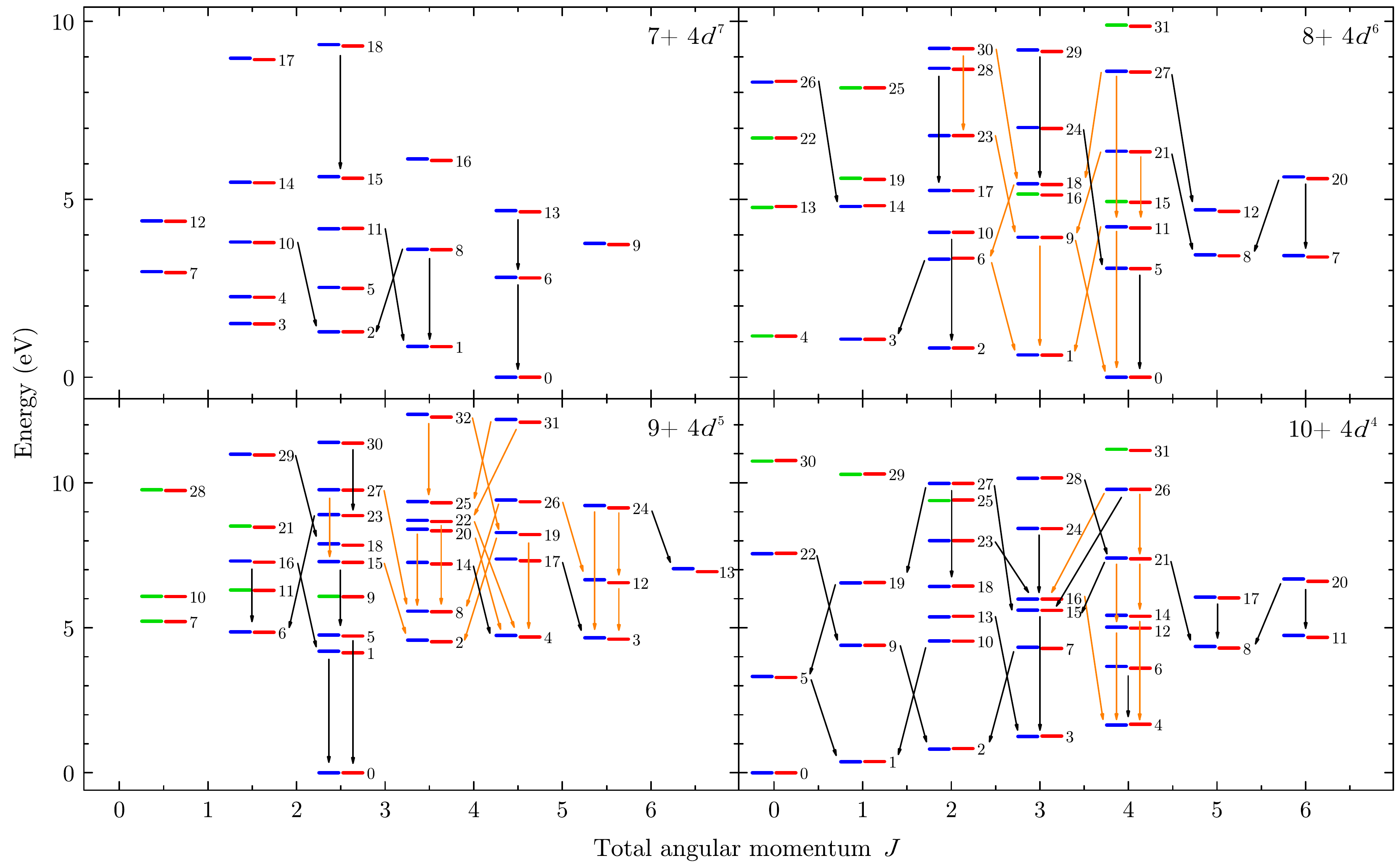}
\caption{Grotrian diagrams for the ions Sn$^{7+}$--Sn$^{10+}$. The energy levels in blue are determined experimentally, for Sn$^{7+}$ they are taken from \cite{Azarov1993}, whereas the Sn$^{8+}$--Sn$^{10+}$ energy levels are results of this work (see Table\,\ref{tab:levels}).  Green levels in Sn$^{8+}$--Sn$^{10+}$ could not be determined experimentally, and thus the values from orthogonal parameters calculations are used. Levels in red are calculated with the \ambit\ code. Arrows indicate experimentally observed optical transitions. Orange arrows indicate transitions which are part of one or more Ritz combinations.
\label{fig:Grotrian}}
\end{figure*}

Table\,\ref{tab:8p9p10p_lines} contains 28 identified lines between the levels of the 4$d^5$ configuration in Sn$^{9+}$. Two lines, at 296.0 and 457.0\,nm wavelength, are doubly classified. Measured intensities of branching ratios (from the ten upper levels numbered 15, 16, 19, 20, 22, 24, 26, 27, 31, and 32 in Table\,\ref{tab:levels}, \emph{cf.} Fig.\,\ref{fig:Grotrian}) are in reasonable agreement with the \textsc{cowan} calculations, except for blended or doubly classified lines (affecting the transitions coupling levels 2-15, 2-19, 4-20, 12-24, and 8-27), and ultraviolet transitions (8-27, 3-24, and 4-24) due to the lower detection efficiency, similar to the identified Sn$^{8+}$ lines. With the identified transitions, 24 levels connected to the ground $^6S_{5/2}$ level were established. Their values, optimized using \textsc{lopt} \cite{Kramida2011}, are listed in Table\,\ref{tab:levels} with their respective uncertainties. The parameter fitting to the available levels resulted in an uncertainty of 41\,cm$^{-1}$ (one-standard-deviation of the fit). As in the case of Sn$^{8+}$, the majority of the levels can be uniquely designated by the largest component of the $LS$-coupling decomposition of the wavefunction. Four levels are named by the second-largest component. Only a single transition, at 399.9\,nm wavelength, is found to be isolated. Thus, as before, the energy for the corresponding lower level $^4F_{5/2}$ was assumed to be equal to the value obtained from the orthogonal parameters method. These two isolated levels were not used in the parameter fitting. In addition to the agreement with the calculations, 15 levels are connected by transitions composing four Ritz combination chains, thus further supporting the identifications.

From the analysis of EUV vacuum spark observations \cite{Churilov2006SnIX--SnXII}, 21 levels of the 4$d^5$ configuration were found as one group not connected to the ground $^6S_{5/2}$ level in the Sn$^{9+}$ spectrum. The uncertainty of these levels relative to the ground level was estimated as several 100\,cm$^{-1}$ common to the whole group, as indicated by the single value $z_5$ in Table\,\ref{tab:levels}. As in Sn$^{8+}$, the comparison between vacuum spark measurements and the current results allows to obtain information on the systematic uncertainty $z_5$. We find that the average value of this systematic shift is $460 \pm 422$\,cm$^{-1}$, the one-standard-deviation spread of which exceeds the current uncertainty of 41\,cm$^{-1}$ by a factor ten, and therefore points to inconsistencies in the previous assignments. From the statistics of the differences $\Delta E_\mathrm{vs}$ (see Table\,\ref{tab:levels}), we identify two groups of levels with common deviations $131 \pm 10$ and $197 \pm 13$\,cm$^{-1}$. These groups comprise four (level numbers 3, 12, 13, and 24) and five levels (numbered 1, 4, 5, 6, and 14), respectively. However, the identification in Ref.\,\cite{Churilov2006SnIX--SnXII} of more than half of the levels in the Sn$^{9+}$ 4$d^5$ configuration presents too large deviations ($\Delta E_\mathrm{vs} > 250$\,cm$^{-1}$) from current values.

\subsubsection*{Spectrum of the Sn$^{10+}$ ion}
The list of 26 identified lines between the levels of the 4$d^4$ configuration is presented in Table\,\ref{tab:8p9p10p_lines}. Three of these (at 297.4, 328.1 and 614.1\,nm wavelength) are doubly classified because they can be ambiguously assigned to two transitions. The measured intensities of the branched transitions (from upper levels 20, 21, 26, and 27 in Table\,\ref{tab:levels}, \emph{cf.} Fig.\,\ref{fig:Grotrian}) are in reasonable agreement with the \textsc{cowan} code calculations except for the same two situations seen in the previous subsections: lines observed in the ultraviolet near the edge of the observable range, affecting the branching ratios related to the upper level 27; blending and double classifications which affect the transitions 15-21, 15-26, 16-26, and 18-27. 

The level energies obtained from the analysis of the experimental spectra belong to two isolated groups. One group consists of 23 levels with the $^5D_1$ level being the lowest in energy. The remaining four levels numbered 2, 7, 9, and 22 form another group. Their relative energy values are optimized using \textsc{lopt} \cite{Kramida2011} and are collected in Table\,\ref{tab:levels}. All of the found levels can be uniquely designated by the largest component of the $LS$-coupling decomposition of the wavefunction. The position of the two groups relative to the ground $^5D_0$ level could not be established from the present identifications. Thus, we assume the spacing between the $^5D_1$ and $^5D_0$ to be equal to the values obtained from the orthogonal parameters method, with a one-standard-deviation uncertainty of 14\,cm$^{-1}$ obtained from the fit. In a similar fashion, the value calculated employing the orthogonal parameters method was assigned to the lowest, $^5D_2$ level of the smaller group. The thus determined energy levels of these groups fall well within statistical uncertainties of the calculated values (\emph{cf.} $\Delta E_{\textrm{orth}}$ in Table\,\ref{tab:levels}). However, they were not used in the fitting procedure to determine the energy parameters shown in Table\,\ref{tab:parameters}. 

The level energies thus obtained in this work can be compared to the levels established in Ref.\,\cite{Churilov2006SnIX--SnXII} as determined from EUV spectra. The levels in that work form four isolated groups, three of which are not connected to the ground level. The uncertainties in relative positions of these three groups were estimated to be several 100\,cm$^{-1}$ \cite{Churilov2006SnIX--SnXII}, parameterized by the values $z_{6,7,8}$ in Table\,\ref{tab:levels}. Analogous to the cases of Sn$^{8+}$ and Sn$^{9+}$, the differences $\Delta E_\mathrm{vs}$ are used to probe the agreement of our results with the previous analysis. We find average values for the shifts $z_6 = 339 \pm 95$\,cm$^{-1}$ and $z_7 = 383 \pm 43$\,cm$^{-1}$, when removing a single outlier in the latter group (level number 14). These one-standard-deviation values are reasonably consistent with the experimental uncertainties. Thus, our data support the identification of five levels with shift $z_6$ and seven levels of the $z_7$ group. The outlier, as well as the levels with shift $z_8$, show much larger discrepancies implying that affected energy levels in Ref.\,\cite{Churilov2006SnIX--SnXII} are called into question.

\subsection{Orthogonal scaling parameters}\label{ssec:parameters}
The orthogonal parameters for the isonuclear sequence Sn$^{6+}$-Sn$^{12+}$ obtained from a fit to the experimental levels are collected in Table\,\ref{tab:parameters}. Here, the orthogonal parameters $O2$, $O2'$, $E_a'$ and $E_b'$ are the orthogonal counterparts of the traditional \textsc{cowan} parameters $F^2$, $F^4$, $\alpha$ and $\beta$ \cite{cowan1981,Windberger2016}. The one-electron magnetic (spin-orbit) operator $\zeta$(4$d$) and the effective three-particle electrostatic operators $T1$ and $T2$ are the same as in the \textsc{cowan} code and ($A_c$-$A_0$) are additional two-particle magnetic parameters. The fitting was performed for the matrices of the interacting 4$d^k$+4$d^{k-1}$5$s$+4$d^{k-2}$5$s^2$ configurations, $k$=8--2 for Sn$^{6+}$--Sn$^{12+}$ respectively. The energy parameters of the unknown 4$d^{k-1}$5$s$+4$d^{k-2}$5$s^2$ configurations therein were fixed at the \textsc{mcdf}-calculated values for the average energies and spin-orbit interactions. The corresponding electrostatic and configuration interaction parameters (the latter ones calculated by the \textsc{cowan} code) were also kept fixed for the 4$d^{k-1}$5$s$+4$d^{k-2}$5$s^2$ configurations, after scaling them by 0.85 from their \emph{ab initio} values. The average energy $E_\mathrm{av}$ is defined such that the ground level energy of the 4$d^k$+4$d^{k-1}$5$s$+4$d^{k-2}$5$s^2$ configurations is equal to zero. The two-particle magnetic parameters ($A_c$-$A_0$) were fixed to the \textsc{mcdf}-calculated values in all instances. For better stability of the fitting parameters, $E_b'$ in Sn$^{11+}$ and $E_a'$ in Sn$^{12+}$ were fixed to the extrapolated values.
\begin{figure}[htb]
\includegraphics[width=8.6cm]{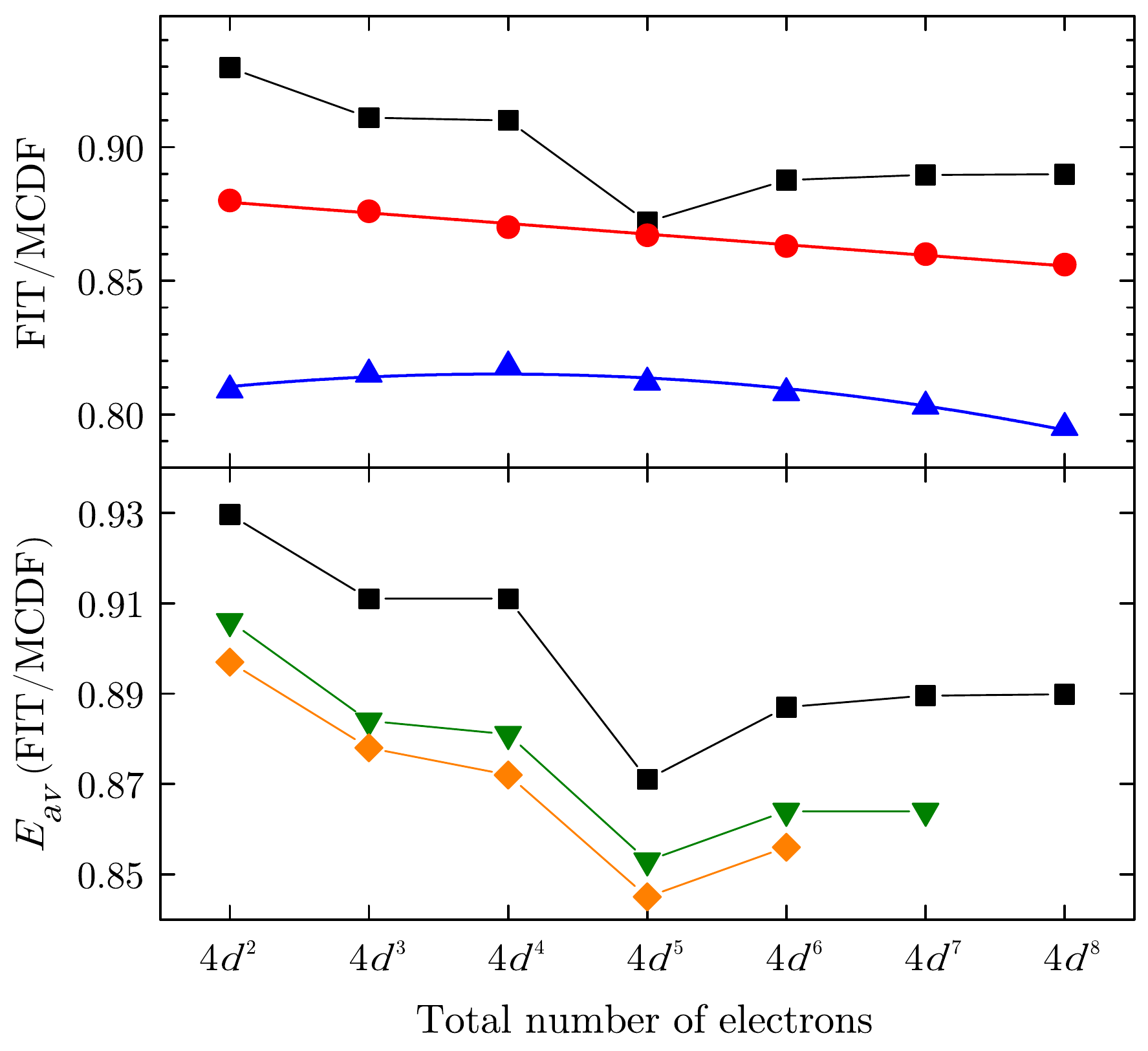}
\caption{(upper) Empirical adjustments of scaling factors compared to the \textsc{mcdf}-calculated values: The ratios (FIT/MCDF) for the electrostatic parameters $O2$ (red circles) and $O2'$ (blue triangles), and for the average energy $E_\mathrm{av}$ (black squares) were obtained by fitting (FIT) to available data. Solid lines represent quadratic fits. (lower) Ratio of the semi-empirical final value (FIT) to the \textsc{mcdf} value for the average energy $E_\mathrm{av}$ along the isonuclear sequences of three elements: black squares Sn$^{12+}$ to Sn$^{6+}$, green inverted triangles Ag$^{9+}$ to Ag$^{4+}$, orange diamonds Pd$^{8+}$ to Pd$^{4+}$.
\label{fig:MCDFscaling_all}}
\end{figure}%

Table\,\ref{tab:parameters} furthermore contains the ratios of the fitted parameters to the parameters obtained from \textsc{mcdf} calculations. Along the Sn$^{6+}$--Sn$^{12+}$ isonuclear sequence the orthogonal energy parameters and the scaling factors can be approximated by linear or weak quadratic dependencies as is visible from Fig.\,\ref{fig:MCDFscaling_all} for the $O2$ and $O2'$ parameters. However, the scaling factor for the average energy presents a discontinuity going from the 4$d^5$ to 4$d^6$ configurations. Fig.\,\ref{fig:MCDFscaling_all} also shows that a similar dependence of the $E_\mathrm{av}$ scaling factors occurs in the Pd and Ag isonuclear sequences. For comparison purposes, the ions Pd$^{4+}$--Pd$^{8+}$ and Ag$^{4+}$--Ag$^{9+}$ have been analyzed using the same orthogonal parameters method here used for Sn. The data were taken from Refs.\,\cite{Raassen1986} (Pd$^{4+}$), \cite{Raassen1987} (Pd$^{5+}$), \cite{Ryabtsev2012} (Pd$^{6+}$), \cite{Ryabtsev2016} (Pd$^{7+}$), \cite{Ryabtsev2011} (Pd$^{8+}$, Ag$^{9+}$), \cite{VanKleef1987} (Ag$^{4+}$), \cite{Joshi1988} (Ag$^{5+}$) and \cite{Ryabtsev2016b} (Ag$^{6+}$--Ag$^{8+}$). The three elements strikingly exhibit the same irregularity in scaling factors of the average energies for the configuration 4$d^5$ (Sn$^{9+}$, Ag$^{6+}$, Pd$^{5+}$), which may be related to the fact that the 4$d^5$ configuration is a half-filled shell. This physical phenomenon, resulting from the maximal exchange contribution in half-filled shells, yields for them a higher binding energy. It is also the cause for the often discussed and somewhat anomalous ground state configurations of the chemical elements Cr ($3d^5\,4s$), Mn ($3d^5\,4s^2$), Mo ($4d^5\,5s$), and Tc ($4d^5\,5s^2$), and analogously for Eu ($4f^7\,5s^2\,p^6\,6s^2$), Gd ($4f^7\,5s^2\,5p^6\,5d^1\,6s^2$), Am ($5f^7\,6s^2\,6p^6\,7s^2$) and Cm ($5f^7\,6s^2\,6p^6\,6d^1\,7s^2$). The \emph{ab initio} \textsc{mcdf} calculations do not accurately account for this exchange contribution, and thus the required empirical correction does not follow a continuous trend, otherwise seen in the filling of the $nd^m$ sub-shell in the isonuclear sequences observed in this experiment and earlier work \cite{Raassen1986,Raassen1987,Ryabtsev2012,Ryabtsev2016,Ryabtsev2011,VanKleef1987,Joshi1988,Ryabtsev2016b} (\emph{cf.} Fig.\,\ref{fig:MCDFscaling_all}). In contrast to this, our CI+MBPT calculations with \ambit\ include this effect from the start.       

\section{Conclusions}
We performed optical and EUV spectroscopy on open 4$d$-shell ions Sn$^{7+}$--Sn$^{10+}$ in a charge-state-resolved manner using an electron beam ion trap and recorded 90 magnetic dipole transitions. Line and level identifications were performed using the semi-empirical orthogonal parameters method and \textsc{cowan} code calculations. Our measurements of transitions in the ground configuration of Sn$^{7+}$ are in good agreement with previous measurements in the EUV \cite{Azarov1993}. The lines and level energies obtained for the 4$d^\textrm{m}$ ($\textrm{m}$=6--4) configurations in Sn$^{8+}$--Sn$^{10+}$ present the most complete data available to date for these ground configurations, with a total of 79 energy levels experimentally determined. Analogous to our recent work on Sn$^{11+}$--Sn$^{14+}$ \cite{Windberger2016}, we conclude that the classification of certain cataloged EUV transitions in previous work \cite{Churilov2006SnIX--SnXII} needs to be revised. Furthermore, these many-valence-electron, open-4$d$ shell ions provide an excellent testing ground for state-of-the-art CI+MBPT calculations, performed with the \ambit\ code. Our \emph{ab initio} calculations are shown to be in very good agreement with our data, validating the predictive power of this theoretical method for these until now challengingly complex electronic systems.

\begin{acknowledgments}
This work is part of and supported by the DFG Collaborative Research Centre ``SFB 1225 (ISOQUANT)''. JB would like to express his gratitude for ARCNL's hospitality during his visit there.
\end{acknowledgments}

\appendix*
\section{Tables}
In this appendix, the tabulated values for all experimentally determined and calculated quantities are presented. Table\,\ref{tab:7plines} collects the measured $M$1 transitions of Sn$^{7+}$. Comparison is made between these transitions, the transitions as inferred from the levels determined in Ref.\,\cite{Azarov1993}, and the transitions predicted by CI+MBPT theory. Moreover, in Table\,\ref{tab:7pambit}, the \ambit-calculated energy levels of the $4d^7$ ground configuration of Sn$^{7+}$ are compared to the level energies determined experimentally in Ref.\,\cite{Azarov1993}. 

Table\,\ref{tab:8p9p10p_lines} shows the wavelengths of the lines measured in the optical domain, along with their identification and values as determined from \ambit\ calculations.

The energy levels of the ground configuration of the ions Sn$^{8+}$--Sn$^{10+}$ are given in Table\,\ref{tab:levels}. Here the level energies optimized with Kramida's \textsc{lopt} \cite{Kramida2010} are shown alongside with the levels calculated with the orthogonal energy parameters method, the results of CI+MBPT calculations, and levels from previous work \cite{Churilov2006SnIX--SnXII}. Finally, the orthogonal parameters used in the semi-empirical calculations are collected in Table\,\ref{tab:parameters}.

\newpage

\begin{table}[H]
\caption{\label{tab:7pambit}
Energy levels of the ground configuration 4$d^7$ for Sn$^{7+}$ (all in cm$^{-1}$). The level energies $E_\mathrm{exp}$ were experimentally determined in Ref.\,\cite{Azarov1993} and are provided along with their approximate $LS$-term. $E_\mathrm{CI+MBPT}$ are energy levels calculated by the \ambit\ code. The difference between the two data sets $\Delta E_{\textsc{CI+MBPT}} = E_{\mathrm{exp}} - E_{\textsc{CI+MBPT}}$ are presented in the last column.}
\begin{ruledtabular}
\begin{tabular}{r c r r r}
\rule[-1.75mm]{0pt}{0mm}Level & Term & $E_{\mathrm{exp}}$ & $E_\mathrm{CI+MBPT}$ & $\Delta E_{\textsc{CI+MBPT}}$ \\
\hline
0  & $^4F_{9/2}$ & 0     & 0     & 0    \\
1  & $^4F_{7/2}$ & 6986  & 6944  & 42   \\
2  & $^4F_{5/2}$ & 10341 & 10318 & 23   \\
3  & $^4F_{3/2}$ & 12153 & 12137 & 16   \\
4  & $^4P_{3/2}$ & 18280 & 18126 & 154  \\
5  & $^4P_{5/2}$ & 20373 & 20123 & 250  \\
6  & $^2G_{9/2}$ & 22636 & 22523 & 113  \\
7  & $^4P_{1/2}$ & 23946 & 23698 & 248  \\
8  & $^2G_{7/2}$ & 29001 & 28924 & 77   \\
9  & $^2H_{11/2}$ & 30312 & 30047 & 265  \\
10 & $^2P_{3/2}$ & 30657 & 30487 & 170  \\
11 & $^2D_{5/2}(3)$ & 33670 & 33762 & -92  \\
12 & $^2P_{1/2}$ & 35458 & 35329 & 129  \\
13 & $^2H_{9/2}$ & 37751 & 37486 & 265  \\
14 & $^2D_{3/2}(3)$ & 44177 & 44051 & 126  \\
15 & $^2F_{5/2}$ & 45452 & 45083 & 369  \\
16 & $^2F_{7/2}$ & 49476 & 49087 & 389  \\
17 & $^2D_{3/2}(1)$ & 73321 & 71994 & 1327 \\
18 & $^2D_{5/2}(1)$ & 75377 & 75089 & 288 
\end{tabular}
\end{ruledtabular}
\end{table}
\begin{table*}[t]
\caption{\label{tab:7plines}
Experimental vacuum wavelengths $\lambda_{\mathrm{exp}}$ and line intensities for Sn$^{7+}$ within its ground electronic configuration [Kr]4$d^7$. Spectra recorded at acceleration potential of 157\,V, which yielded maximum fluorescence signal. Intensity integrals from  Gaussian fits were corrected for the grating efficiency. Wavelengths $\lambda_{\mathrm{Ritz}}$ are determined from the energy levels of Sn$^{7+}$ given in Ref.\,\cite{Azarov1993}. Transition probabilities $gA_{ij,\textsc{cowan}}$ are calculated with the \textsc{cowan} code based on those levels. Wavelengths $\lambda_{\textsc{CI+MBPT}}$ are \emph{ab initio} results from our CI+MBPT calculations. ``Transition'' column shows lower and upper levels as used in Fig.\,\ref{fig:Grotrian}. Approximate $LS$-terms are given in the last column. Numbers in brackets are sequential indices as defined by Nielson and Koster \cite{nielson1963} to differentiate levels with the same $LSJ$ values. Superscripts $bl$ mark spectral blends.}
\begin{ruledtabular}
\begin{tabular}{l r r r r c c}
\multicolumn{1}{c}{$\lambda_{\mathrm{exp}}$}	& \multicolumn{1}{c}{Intensity \,\,} & \multicolumn{1}{c}{$\lambda_{\mathrm{Ritz}}$}& \multicolumn{1}{c}{$gA_{ij,\textsc{cowan}}$} & \multicolumn{1}{c}{$\lambda_{\textsc{CI+MBPT}}$}& Transition & Terms \\
 \multicolumn{1}{c}{\rule[-1.75mm]{0pt}{0mm}(nm)}&
    \multicolumn{1}{c}{(arb. units)}& \multicolumn{1}{c}{(nm)}&
    \multicolumn{1}{c}{(s$^{-1}$)} & \multicolumn{1}{c}{(nm)} & (see Fig.\,\ref{fig:Grotrian})& \\
\hline	
 333.9$^{bl}$	& 132  &  334.2  & 167  & 333.3   & 15-18 &   $^2F_{5/2}$-$^2D_{5/2}(1)$		  \\
 374.5	& 271  &  374.8  & 375  & 372.9   & 1-11  &        $^4F_{7/2}$-$^2D_{5/2}(2)$		  \\								
 386.0	& 68   &  386.1  & 57   & 384.6   & 16-18 &         $^2F_{7/2}$-$^2D_{5/2}(1)$		  \\
 441.9	& 142  &  441.8  & 262  & 444.0   & 0-6   &      $^4F_{9/2}$-$^2G_{9/2}$		  \\
 454.3	& 12   &  454.2  & 66   & 455.0    & 1-8   &     $^4F_{7/2}$-$^2G_{7/2}$		  \\
 492.2	& 69   &  492.2  & 44   & 495.8    & 2-10  &     $^4F_{5/2}$-$^4P_{3/2}$		  \\
 536.0$^{bl}$	& 35   &  535.9  & 32   & 537.5    & 2-8   &   $^4F_{5/2}$-$^2G_{7/2}$		  \\
 660.9	& 116  &  661.6  & 170  & 668.3   & 6-13  &        $^2G_{9/2}$-$^2H_{9/2}$\\
\end{tabular}
\end{ruledtabular}
\end{table*}
\newpage
\mbox{}
\newpage
\mbox{}
\newpage
\begin{longtable*}{l c l r r r r c c c}
\caption{Vacuum wavelengths and line intensities for Sn$^{8+}$--Sn$^{10+}$ ions at the acceleration potential maximizing ion fluorescence. Intensities are taken from Gaussian fits and corrected for the grating efficiency. Wavelengths $\lambda_\mathrm{orth}$ are calculated from level energies from Table\,\ref{tab:levels}. Transition probabilities $g A_{ij,\textsc{cowan}}$ are determined with the \textsc{cowan} code using the same level energies. Wavelengths $\lambda_{\textsc{CI+MBPT}}$ calculated \emph{ab initio} with the \ambit\ CI+MBPT code. ``Transition'' refers to levels shown in Fig.\,\ref{fig:Grotrian}. Configurations and approximate $LS$-coupling terms are given in the last two columns (the numbers in brackets are sequential indices \cite{nielson1963} for distinction of levels with the same $LSJ$ values). The superscript $bl$ indicates line blends, and the superscript $D$ marks doubly classified lines (i.e.~which can be ambiguously assigned to two transitions). \label{tab:8p9p10p_lines}}\\
\hline\hline
\multicolumn{1}{c}{\rule{0pt}{3mm}Ion \, \, \, } 					& \multicolumn{1}{c}{V$_{\textrm{max}}$} 	& \multicolumn{1}{c}{$\lambda_{\mathrm{exp}}$}	& \multicolumn{1}{c}{Intensity \, \,} & \multicolumn{1}{c}{$\lambda_{\mathrm{orth}}$}& \multicolumn{1}{c}{$gA_{ij,\textsc{cowan}}$} & {$\lambda_{\textsc{CI+MBPT}}$} & Transition & Configuration & Term symbol \\
		&  \multicolumn{1}{c}{\rule[-1.5mm]{0pt}{0mm}(V)}	& \multicolumn{1}{c}{(nm)}&
    \multicolumn{1}{c}{(arb. units)}& \multicolumn{1}{c}{(nm)}&
    \multicolumn{1}{c}{(s$^{-1}$)} & \multicolumn{1}{c}{(nm)}& (see Fig.\,\ref{fig:Grotrian})& & \\
\hline
\endfirsthead
\caption[]{(continued)}\\
\hline\hline
\multicolumn{1}{c}{\rule{0pt}{3mm}Ion \, \, \, } 					&
\multicolumn{1}{c}{V$_{\textrm{max}}$} 	&
\multicolumn{1}{c}{$\lambda_{\mathrm{exp}}$}	& \multicolumn{1}{c}{Intensity \,
\,} & \multicolumn{1}{c}{$\lambda_{\mathrm{orth}}$}&
\multicolumn{1}{c}{$gA_{ij,\textsc{cowan}}$} & \multicolumn{1}{c}
{$\lambda_{\textsc{CI+MBPT}}$}& Transition & Configuration & Term symbol \\
		&  \multicolumn{1}{c}{\rule[-1.5mm]{0pt}{0mm}(V)}	& \multicolumn{1}{c}{(nm)}&
    \multicolumn{1}{c}{(arb. units)}& \multicolumn{1}{c}{(nm)}&
    \multicolumn{1}{c}{(s$^{-1}$)} & \multicolumn{1}{c}{(nm)}& (see Fig.\,\ref{fig:Grotrian})& & \\
\hline
\endhead
\hline
\endfoot
\hline\hline
\endlastfoot
\rule{0pt}{3.25mm}8+    	& 137   & 283.4 				& 18    & 284   & 104 &283.1 & 11-27   &   [Kr]4$d^6$	  & $^3F_{4}(2)$-$^3F_{4}(1)$		  \\
      	&       & 293.3 				& 42    & 293   & 226 &295.2 & 0-11    &          		    & $^5D_{4}$-$^3F_{4}(2)$		  \\
      	&       & 313.5 				& 17    & 314   & 54  &315.1 & 5-24    &            		  & $^3H_{4}$-$^1F_{3}$		  \\
				&       & 315.0 				& 16    & 315   & 60  &315.4 & 0-9     &            	  	& $^5D_{4}$-$^3F_{3}(2)$		  \\
				&       & 317.6 				& 45    & 318   & 192 &316.6 & 12-27   &            		  & $^3G_{5}$-$^3F_{4}(1)$		  \\
				&       & 326.0 				& 15    & 326   & 105 &325.2 & 18-30   &            		  & $^3D_{3}$-$^3P_{2}(1)$		  \\
				&       & 330.4 				& 70    & 330   & 279 &331.6 & 18-29   &            		  & $^3D_{3}$-$^3F_{3}(1)$		  \\				
				&       & 344.6 				& 25    & 344   & 31  &346.5 & 1-11    &             		& $^5D_{3}$-$^3F_{4}(2)$		  \\
				&       & 354.8 				& 33    & 355   & 102 &352.3 & 14-26   &            	 	  & $^3P_{1}(2)$-$^3P_{0}(1)$		  \\
				&       & 360.2$^{bl}$	& 43    & 360   & 120 &363.5 & 17-28   &            		  & $^3D_{2}$-$^3F_{2}(1)$		  \\
				&       & 374.5$^{bl}$ 	& 271   & 374   & 131 &374.6 & 1-9     &            		  & $^5D_{3}$-$^3F_{3}(2)$		  \\
				&       & 381.2 				& 36    & 381   & 65  &381.3 & 2-10    &            		  & $^5D_{2}$-$^3F_{2}(2)$		  \\
				&       & 392.1 				& 24    & 392   & 78  &392.2 & 18-27   &            		  & $^3D_{3}$-$^3F_{4}(1)$		  \\				
				&       & 404.6 				& 60    & 404   & 239 &405.7 & 0-5     &            		  & $^5D_{4}$-$^3H_{4}$		  \\
				&       & 426.6 				& 54    & 428   & 186 &424.0 & 8-21    &            		  & $^3H_{5}$-$^1G_{4}(2)$		  \\
				&       & 434.1 				& 36    & 434   & 107 &434.2 & 9-23    &            		  & $^3F_{3}(2)$-$^1D_{2}(2)$		  \\
				&       & 460.9 				& 138   & 461   & 315 &454.5 & 1-6     &            		  & $^5D_{3}$-$^3P_{2}(2)$		  \\
				&       & 505.7 				& 38    & 505   & 107 &507.3 & 23-30   &            		  & $^1D_{2}(2)$-$^3P_{2}(1)$		  \\
				&       & 513.2$^{bl}$	& 85    & 515   & 89  &516.1 & 9-21    &            		  & $^3F_{3}(2)$-$^1G_{4}(2)$		  \\
				&       & 551.8 				& 40    & 552   & 65  &542.9 & 3-6     &            	  	& $^5D_{1}$-$^3P_{2}(2)$		  \\
				&       & 560.9 				& 149   & 560   & 185 &568.9 & 7-20    &            		  & $^3H_{6}$-$^1I_{6}$		  \\
				&       & 566.8 				& 91    & 565   & 87  &576.5 & 8-20    &            		  & $^3H_{5}$-$^1I_{6}$		  \\
				&       & 584.4$^{D}$		& 118   & 585   & 62  &599.1 & 6-18    &            		  & $^3P_{2}(2)$-$^3D_{3}$		  \\\vspace{2mm}
				&       & 584.4$^{D}$		& 118   & 585   & 291 &581.0 & 11-21   &             		& $^3F_{4}(2)$-$^1G_{4}(2)$		  \\
9+    	& 137   & 261.0 				& 66    & 261   & 304 &262.9 & 0-5     &   [Kr]4$d^5$	  & $^6S_{5/2}$-$^4G_{5/2}$		  \\
      	&       & 272.0 				& 53    & 272   & 418 &273.8 & 3-24    &   	          	& $^4G_{11/2}$-$^2H_{11/2}$		\\
        &       & 276.6 				& 30    & 276   & 209 &278.4   & 4-24    &            		  & $^4G_{9/2}$-$^2H_{11/2}$		\\
				&       & 296.0$^{D}$		& 46    & 296   & 115 &295.2   & 8-27    &            		  & $^4D_{7/2}$-$^2F_{5/2}(1)$		  \\
				&       & 296.0$^{D}$  	& 46    & 296   & 149 &299.6   & 0-1     &            		  & $^6S_{5/2}$-$^4G_{5/2}$		  \\
				&       & 304.8 				& 25    & 303   & 69  &306.2    & 19-32   &        		      & $^2H_{9/2}$-$^2G_{7/2}(1)$		  \\
				&       & 306.6$^{bl}$	& 34    & 306   & 16  &307.7    & 6-23    &            		  & $^4P_{3/2}$-$^2F_{5/2}(2)$		  \\
				&       & 312.1 				& 62    & 312   & 151 &311.5   & 4-22    &            	   	& $^4G_{9/2}$-$^2F_{7/2}(1)$		  \\
				&       & 323.9 				& 50    & 324   & 105 &326.7   & 8-26    &            		  & $^4D_{7/2}$-$^2G_{9/2}(2)$		  \\
				&       & 333.9$^{bl}$  & 70    & 333   & 132 &335.0   & 2-19    &            		  & $^4G_{7/2}$-$^2H_{9/2}$		  \\
				&       & 338.7$^{bl}$	& 219   & 339   & 222 &338.2   & 4-20    &            		  & $^4G_{9/2}$-$^2G_{7/2}(2)$		  \\
				&       & 349.1$^{bl}$	& 271   & 349   & 297 &350.7   & 4-19    &            		  & $^4G_{9/2}$-$^2H_{9/2}$		  \\
				&       & 356.8 				& 41    & 354   & 57  &362.0    & 22-31   &            		  & $^2F_{7/2}(1)$-$^2G_{9/2}(1)$		  \\
				&       & 395.5 				& 64    & 394   & 131 &398.1   & 8-22    &            		  & $^4D_{7/2}$-$^2F_{7/2}(1)$		  \\
				&       & 398.3 				& 68    & 398   & 86  &396.7    & 1-16    &            		  & $^4G_{5/2}$-$^4F_{3/2}$		  \\
				&       & 399.9 				& 26    & 401   & 42  &398.2    & 18-29   &            		  & $^4F_{5/2}$-$^2D_{3/2}(2)$		  \\
				&       & 413.5 				& 43    & 415   & 133 &419.2   & 25-32   &            		  & $^2F_{7/2}(2)$-$^2G_{7/2}(1)$		  \\
				&       & 438.9 				& 46    & 439   & 155 &442.7   & 8-20    &            	  	& $^4D_{7/2}$-$^2G_{7/2}(2)$		  \\
				&       & 439.7 				& 34    & 441   & 109 &446.2   & 25-31   &            		  & $^2F_{7/2}(2)$-$^2G_{9/2}(1)$		  \\
				&       & 452.4 				& 32    & 453   & 105 &444.8   & 12-26   &            		  & $^2I_{11/2}$-$^2G_{9/2}(2)$		\\
				&       & 457.0$^{D}$ 	& 58    & 456   & 99  &453.4    & 2-15    &            		  & $^4G_{7/2}$-$^2D_{5/2}(3)$		  \\
				&       & 457.0$^{D}$ 	& 58    & 456   & 82  &458.0    & 3-17    &        		      & $^4G_{11/2}$-$^4F_{9/2}$	  \\
				&       & 485.0$^{bl}$	& 358   & 484   & 472 &480.5   & 12-24   &       		      & $^2I_{11/2}$-$^2H_{11/2}$		\\
				&       & 489.2 				& 414   & 489   & 562 &489.1   & 5-15    &            	  	& $^4G_{5/2}$-$^2D_{5/2}(3)$		  \\
				&       & 492.2 				& 107   & 492   & 69  &491.9    & 4-14    &              		& $^4G_{9/2}$-$^4F_{7/2}$		  \\
				&       & 497.0 				& 55    & 496   & 107 &497.3   & 23-30   &            		  & $^2F_{5/2}(2)$-$^2D_{5/2}(2)$		  \\
				&       & 501.4 				& 167   & 502   & 339 &496.6   & 15-27   &            		  & $^2D_{5/2}(3)$-$^2F_{5/2}(1)$		  \\
				&       & 507.6 				& 105   & 508   & 103 &511.4   & 6-16    &            		  & $^4P_{3/2}$-$^4F_{3/2}$		  \\
				&       & \rule[-1.5mm]{0pt}{0mm}570.8 				& 76    & 569   & 151 &563.8   & 13-24   &            		  & $^2I_{13/2}$-$^2H_{11/2}$		\\\vspace{2mm}
				&\rule{0pt}{3.25mm}& 618.9 				& 137   & 621   & 119 &636.2   & 3-12    &            	   	& $^4G_{11/2}$-$^2I_{11/2}$		\\	
10+    	& 217   & 283.7 				& 13   & 283   & 278  &283.0   & 15-27    &   [Kr]4$d^4$		& $^3F_{3}(2)$-$^3F_{2}(1)$		  \\
        &       & 284.7 				& 18   & 285   & 114  &286.3   & 3-15     &            		  & $^5D_{3}$-$^3F_{3}(2)$		  \\
				&       & 286.0 				& 33   & 286   & 163  &287.4   & 4-16     &            		  & $^5D_{4}$-$^3D_{3}$		  \\
				&       & 297.4$^{D}$ 	& 15   & 297   & 110  &298.8   & 1-10     &            		  & $^5D_{1}$-$^3F_{2}(2)$		  \\
				&       & 297.4$^{D}$		& 15   & 298   & 91   &296.7    & 15-26    &            		  & $^3F_{3}(2)$-$^3F_{4}(1)$		  \\
				&       & 300.5 				& 15   & 300   & 73   &300.2    & 3-13     &            		  & $^5D_{3}$-$^3P_{2}(2)$		  \\
				&       & 328.1$^{D}$ 	& 112  & 328   & 137  &333.1   & 4-14     &             	  & $^5D_{4}$-$^3G_{4}$		  \\
				&       & 328.1$^{D}$		& 112  & 328   & 261  &327.4   & 16-26    &            		  & $^3D_{3}$-$^3F_{4}(1)$		  \\
				&       & 346.8 				& 91   & 347   & 278  &348.2   & 2-9      &            		  & $^5D_{2}$-$^3P_{1}(2)$		  \\
				&       & 349.1$^{bl}$	& 271  & 349   & 93   &349.7    & 18-27    &            		  & $^3D_{2}$-$^3F_{2}(1)$		  \\
				&       & 353.0 				& 38   & 353   & 82   &359.0    & 2-7      &            		  & $^5D_{2}$-$^3G_{3}$		  \\
				&       & 361.9 				& 14   & 361   & 95   &363.1    & 19-27    &            		  & $^3D_{1}$-$^3F_{2}(1)$		  \\
				&       & 367.7 				& 145  & 368   & 227  &374.6   & 4-12     &            		  & $^5D_{4}$-$^3F_{4}(2)$		  \\
				&       & 383.7 				& 28   & 384   & 84   &378.0    & 5-19     &            		  & $^3P_{0}(2)$-$^3D_{1}$		  \\
				&       & 392.7 				& 68   & 392   & 124  &390.6   & 9-22     &            		  & $^3P_{1}(2)$-$^1S_{0}(2)$		  \\
				&       & 407.4 				& 38   & 408   & 87   &402.4    & 8-21     &            		  & $^3H_{5}$-$^1G_{4}(2)$		  \\
				&       & 421.8 				& 26   & 421   & 139  &428.3   & 1-5      &            		  & $^5D_{1}$-$^3P_{0}(2)$		  \\
				&       & 450.5 				& 18   & 450   & 88   &444.7    & 21-28    &            		  & $^1G_{4}(2)$-$^3F_{3}(1)$		  \\
				&       & 508.2 				& 127  & 508   & 193  &510.3   & 16-24    &            		  & $^3D_{3}$-$^1F_{3}$		  \\
				&       & 520.7 				& 84   & 521   & 310  &518.6   & 12-21    &            		  & $^3F_{4}(2)$-$^1G_{4}(2)$		  \\
				&       & 524.2 				& 17   & 525   & 45   &516.5    & 21-26    &            		  & $^1G_{4}(2)$-$^3F_{4}(1)$		  \\
				&       & 534.4 				& 117  & 535   & 234  &538.4   & 8-20     &            		  & $^3H_{5}$-$^1I_{6}$		  \\
				&       & 614.1$^{D}$ 	& 95   & 614   & 102  &643.9   & 4-6      &            		  & $^5D_{4}$-$^3H_{4}$		  \\
				&       & 614.1$^{D}$ 	& 95   & 616   & 106  &616.1   & 16-23    &            		  & $^3D_{3}$-$^1D_{2}(2)$		  \\
				&       & 628.3 				& 69   & 630   & 116  &626.8   & 14-21    &            		  & $^3G_{4}$-$^1G_{4}(2)$		  \\
				&       & 639.9 				& 157  & 642   & 201  &642.9   & 11-20    &            		  & $^3H_{6}$-$^1I_{6}$		  \\
				&       & 689.5$^{bl}$	& 85   & 690   & 164  &697.0   & 15-21    &            		  & $^3F_{3}(2)$-$^1G_{4}(2)$		  \\			
				&       & 728.1 				& 39   & 727   & 103  &713.1   & 8-17     & \rule[-1mm]{0pt}{0mm} & $^3H_{5}$-$^3G_{5}$		  \\
\end{longtable*}
\newpage
\mbox{}
\newpage
\mbox{}
\newpage
\begin{longtable*}{c r c r r r r r r r r r r}
\caption{Energy levels of the Sn$^{8+}$ 4$d^6$, Sn$^{9+}$ 4$d^5$, and Sn$^{10+}$ 4$d^4$ configurations (in cm$^{-1}$) adjusted with the \textsc{lopt} algorithm \cite{Kramida2011} based on the measured transitions. Levels labels use approximate $LS$-coupling terms. Numbers in brackets display sequential indices \cite{nielson1963} to differentiate levels having the same $LSJ$ values. Uncertainties $x_j$ ($j=1-5$) and $y$ are given as the one-standard-deviation of the orthogonal parameters fit for the respective configuration: $x_{1,2,3}=\pm 16\,\mathrm{cm}^{-1}$; $x_4=\pm 41\,\mathrm{cm}^{-1}$; $x_5= y =\pm 14\,\mathrm {cm}^{-1}$. The dispersive energy uncertainty $D_1$ is close to the minimum uncertainty of separation from other levels, and the energy uncertainty $D_2$ is that relative to the ground level of the configuration (\emph{cf.} \cite{Kramida2011}). $N$ is the total number of lines connected to the level. $E_{\mathrm{orth}}$ values are semi-empirical energy levels calculated with the orthogonal parameters in Table\,\ref{tab:parameters}. The $E_{\textsc{CI+MBPT}}$ values are \emph{ab initio} energy levels calculated using the \ambit\ CI+MBPT code. Differences between experimental and calculated values appear in columns $\Delta E_{\mathrm{orth}}$ ($E_{\mathrm{exp}}-E_{\mathrm{orth}}$) and $\Delta E_{\textsc{CI+MBPT}}$ ($E_{\mathrm{exp}} - E_{\textsc{CI+MBPT}}$). Energies determined from previous vacuum spark measurement \cite{Churilov2006SnIX--SnXII} shown as $E_{\mathrm{vs}}$; $\Delta E_{\mathrm{vs}} = E_{\mathrm{exp}} - E_{\mathrm{vs}}$, their deviations. Uncertainties in the systematic common shifts of the identified level groups $z_i$ ($i=1-8$) \cite{Churilov2006SnIX--SnXII} are of the order of several hundreds of cm$^{-1}$ (see main text). The uncertainty of the relative level energies within each of these groups was estimated at 10\,cm$^{-1}$. \label{tab:levels}} \\
\hline\hline
\rule{0pt}{3mm}Ion & \rule[-1mm]{0pt}{0mm}Level & Term & $E_{\mathrm{exp}}$ & $D_1$ & $D_2$ & $N$ &
$E_{\mathrm{orth}}$ & $\Delta E_{\mathrm{orth}}$ & $E_{\textsc{CI+MBPT}}$ & $\Delta E_{\textsc{CI+MBPT}}$ & $E_{\mathrm{vs}}$ & $\Delta E_{\mathrm{vs}}$ \\
\hline
\endfirsthead
\caption[]{(continued)}\\
\hline\hline
\rule{0pt}{3mm}Ion & Level & Term & $E_{\mathrm{exp}}$ & $D_1$ & $D_2$ & $N$ &
$E_{\mathrm{orth}}$ & $\Delta E_{\mathrm{orth}}$ & $E_{\textsc{CI+MBPT}}$ & $\Delta E_{\textsc{CI+MBPT}}$ & $E_{\mathrm{vs}}$ & $\Delta E_{\mathrm{vs}}$ \\
\hline
\endhead
\hline
\endfoot
\hline\hline
\endlastfoot
\rule{0pt}{3.25mm}8+ $4d^6$ & 0  & $^5D_4$  & 0        & 30 & 0  & 2 & -5 & 5 & 0  & 0 & 0 & 0 \\
 & 1  & $^5D_3$  & 5\,075     & 13 & 30 & 3 & 5\,064 & 11 & 5\,011 & 64 & 5\,050 & 25\\
 & 2  & $^5D_2$  & $6\,634+x_2$  & 0  & 0  & 0 & 6\,634 & 0 & 6\,626 & 8& $6\,670+z_1$ & -36 \\
 & 3  & $^5D_1$  & 8\,648     & 13 & 40 & 1 & 8\,636 & 12 & 8\,593 & 55 &$8\,670+z_1$ & -22 \\
 & 4  & $^5D_0$  &          &    &    &   & 9\,345 & &9\,307& & & \\
 & 5  & $^3H_4$  & 24\,716    & 24 & 24 & 1 & 24\,726 & -10 &24\,651 & 65 & $24\,685+z_2$ & 31 \\
 & 6  & $^3P_{2}(2)$  & 26\,785    & 16 & 40 & 2 & 26\,771 & 14 &27\,011 & -226 & & \\
 & 7  & $^3H_6$  & 27\,592    & 13 & 43 & 1 & 27\,604 & -12 &27\,270 & 322 & $27\,610+z_2$ & -18 \\
 & 8  & $^3H_5$  & 27\,778    & 22 & 39 & 1 & 27\,781 & -3 &27\,503 & 275 & $27\,710+z_2$ & 68 \\
 & 9  & $^3F_{3}(2)$  & 31\,740    & 12 & 30 & 4 & 31\,736 & 4 &31\,709 & 31 & $31\,747+z_3$ & -7 \\
 & 10 & $^3F_{2}(2)$  & $32\,847+x_2$ & 28 & 28 & 1 & 32\,847 & 0 &32\,855 & -8 & $33\,028+z_3$ & -181 \\
 & 11 & $^3F_{4}(2)$  & 34\,103    & 11 & 30 & 4 & 34\,102 & 1 &33\,873 & 230 & $34\,220+z_2$ & -117 \\
 & 12 & $^3G_5$  & 37\,908    & 40 & 59 & 1 & 37\,930 & -22 &37\,616 & 292 & $37\,950+z_2$ & -42 \\
 & 13 & $^1S_{0}(4)$  &          &    &    &   & 38\,532& &38\,684 & & &  \\
 & 14 & $^3P_{1}(2)$  & $38\,694+x_1$ & 0  & 0  & 0 & 38\,694 & 0 &38\,903 & -209 & & \\
 & 15 & $^3G_4$  &          &    &    &   & 39\,872 & &39\,674 & & $39\,609+z_2$ & \\
 & 16 & $^3G_3$  &          &    &    &   & 41\,548 & &41\,310 &  & & \\
 & 17 & $^3D_2$  & $42\,340+x_3$ & 0  & 0  & 0 & 42\,340 & 0 &42\,287 & 53 & $41\,787+z_2$ & 553 \\
 & 18 & $^3D_3$  & 43\,879    & 17 & 40 & 3 & 43\,887 & -8 &43\,704 & 175 & & \\
 & 19 & $^3D_1$  &          &    &    &   & 45\,061&  &44\,847 & & & \\
 & 20 & $^1I_6$  & 45\,421    & 13 & 41 & 1 & 45\,399 & 22 &45\,032 & 389 & $45\,440+z_2$ & -19 \\
 & 21 & $^1G_{4}(2)$  & 51\,219    & 11 & 30 & 2 & 51\,217 & 2 &51\,085 & 134 & $50\,840+z_4$ & 379 \\
 & 22  & $^3P_{0}(2)$  &          &    &    &   & 54\,202 & &54\,250  & & & \\
 & 23 & $^1D_{2}(2)$  & 54\,777    & 13 & 40 & 2 & 54\,795 & -18 &54\,742 & 35 & & \\
 & 24 & $^1F_3$  & 56\,613    & 41 & 47 & 1 & 56\,586 & 27 &56\,385 & 228 & & \\
 & 25 & $^3P_{1}(1)$  &          &    &    &   & 65\,561 & &65\,611 & & & \\
 & 26 & $^3P_{0}(1)$  & $66\,874+x_1$ & 32 & 32 & 1 & 66\,875 & -1 &67\,067 & -193 & & \\
 & 27 & $^3F_{4}(1)$  & 69\,394    & 23 & 40 & 2 & 69\,401 & -7 &69\,198 & 196 & $68\,566+z_2$ & 828 \\
 & 28 & $^3F_{2}(1)$  & $70\,006+x_3$ & 31 & 31 & 1 & 70\,006 & 0 &69\,800 & 206 & & \\
 & 29 & $^3F_{3}(1)$  & 74\,146    & 37 & 54 & 1 & 74\,144 & 2 &73\,860 & 286 &$73\,385+z_2$ & 761 \\
 & 30 & $^3P_{2}(1)$  & 74\,552    & 14 & 40 & 2 & 74\,548 & 4 &74\,454 & 98 & & \\
 & 31 & $^1G_{4}(1)$  &          &    &    &   & 79\,767 & &79\,565 & & $79\,186+z_2$ & \\
 & 32 & $^1D_{2}(1)$  &          &    &    &   & 101\,675 &  &101\,319 & & $99\,838+z_4$ &\\\vspace{2mm}
 & 33 & $^1S_{0}(1)$  &          &    &    &   & 131\,838 & &131\,874 & & $130\,008 + z_4$ & \\
9+ $4d^5$ & 0  & $^6S_{5/2}$ & 0   & 59 & 0 & 1 & -17 & 17 & 0 & 0 & 0 & 0 \\
 & 1  & $^4G_{5/2}$ & 33\,784   & 61 & 61 & 1 & 33\,748 & 36 &33\,381 & 403 & $33\,582+z_5$ & 202 \\
 & 2  & $^4G_{7/2}$ & 36\,874   & 21 & 70 & 2 & 36\,834 & 40 &36\,421 & 453 & $36\,610+z_5$ & 264 \\
 & 3  & $^4G_{11/2}$ & 37\,535   & 10 & 90 & 2 & 37\,576 & -41 &37\,145 & 390 & $37\,399+z_5$ & 136 \\
 & 4  & $^4G_{9/2}$ & 38\,170   & 20 & 80 & 4 & 38\,173 & -3 &37\,759 & 411 & $37\,958+z_5$& 212 \\
 & 5  & $^4G_{5/2}$ & 38\,315   & 16 & 59 & 1 & 38\,282 & 33 &38\,032 & 283 & $38\,110+z_5$ & 205 \\
 & 6  & $^4P_{3/2}$ & 39\,190   & 16 & 68 & 1 & 39\,183 & 7 &39\,035 & 155 & $39\,010+z_5$ & 180 \\
 & 7  & $^4P_{1/2}$ &         &  &  &  & 42\,159 & &42\,060 & & & \\
 & 8  & $^4D_{7/2}$ & 44\,915   & 15 & 80 & 4 & 44\,958 & -43 &44\,737 & 178 & $44\,470+z_5$ & 445 \\
 & 9  & $^4D_{5/2}$ &         &  &  &  & 49\,065 & &48\,907& & & \\
 & 10 & $^4D_{1/2}$ &         &  &  &  & 49\,104 & &48\,996& & & \\
 & 11 & $^4D_{3/2}$ &         &  &  &  & 50\,861 & &50\,753& & & \\
 & 12 & $^2I_{11/2}$ & 53\,692   & 8 & 80 & 3 & 53\,685 & 7 &52\,863 & 829 & $53\,554+z_5$ & 138 \\
 & 13 & $^2I_{13/2}$ & 56\,792   & 12 & 83 & 1 & 56\,765 & 27 &55\,937 &855& $56\,660+z_5$ & 132 \\
 & 14 & $^4F_{7/2}$ & 58\,487   & 16 & 77 & 1 & 58\,491 & -4 &58\,088 & 399 & $58\,300+z_5$ & 187 \\
 & 15 & $^2D_{5/2}(3)$ & 58\,756   & 14 & 60 & 2 & 58\,721 & 35 &58\,479 & 277 & $58\,370+z_5$ & 386 \\
 & 16 & $^4F_{3/2}$ & 58\,891   & 25 & 66 & 1 & 58\,848 & 43 &58\,588 & 303 & & \\
 & 17 & $^4F_{9/2}$ & 59\,417   & 28 & 87 & 1 & 59\,469 & -52 &58\,981 & 436 & $58\,850+z_5$ & 567 \\
 & \rule[-1.5mm]{0pt}{0mm}18 & $^4F_{5/2}$ & $63\,643+x_4$ & 0 & 0 & 0 & 63\,643 & 0 &63\,291 & 352 & & \\
 & \rule{0pt}{3.25mm}19 & $^2H_{9/2}$ & 66\,824   & 22 & 70 & 3 & 66\,846 & -22 &66\,273 & 551 & $66\,427+z_5$ & 397 \\
 & 20 & $^2G_{7/2}(2)$ & 67\,698   & 18 & 80 & 2 & 67\,687 & 11 &67\,325 & 373 & $66\,975+z_5$ & 723 \\
 & 21 & $^2D_{3/2}(3)$ &         &  &  &  & 68\,584&  &68\,321&  & & \\
 & 22 & $^2F_{7/2}(1)$ & 70\,199   & 18 & 80 & 3 & 70\,228 & -29 &69\,859 & 340 & $70\,185+z_5$ & 14 \\
 & 23 & $^2F_{5/2}(2)$ & 71\,806   & 43 & 80 & 1 & 71\,837 & -31 &71\,529 & 277 & & \\
 & 24 & $^2H_{11/2}$ & 74\,311   & 16 & 80 & 3 & 74\,338 &-27 &73\,674 & 637 & $74\,195+z_5$ & 116 \\
 & 25 & $^2F_{7/2}(2)$ & 75\,470   & 16 & 80 & 2 & 75\,423 & 47 &75\,073 & 397 & $74\,385+z_5$ & 1\,085 \\
 & 26 & $^2G_{9/2}(2)$ & 75\,795   & 18 & 80 & 2 & 75\,816 & -21 &75\,347 & 448 & $75\,345+z_5$ & 450 \\
 & 27 & $^2F_{5/2}(1)$ & 78\,700   & 16 & 60 & 2 & 78\,654 & 46 &78\,616 & 84 & & \\
 & 28 & $^2S_{1/2}$ &         &  &  &  & 78\,719&  &78\,457& & & \\
 & 29 & $^2D_{3/2}(2)$ & $88\,649+x_4$ & 25 & 25 & 1 & 88\,702 & -53 &88\,405 & 244 & & \\
 & 30 & $^2D_{5/2}(2)$ & 91\,927   & 16 & 81 & 1 & 91\,976 & -49 &91\,640 & 287 & $90\,911+z_5$ & 1\,016 \\
 & 31 & $^2G_{9/2}(1)$ & 98\,217   & 18 & 80 & 2 & 98\,228 & -11 &97\,485 & 732 & $96\,800+z_5$ & 1\,417 \\
 & 32 & $^2G_{7/2}(1)$ & 99\,649   & 21 & 80 & 2 & 99\,568 & 81 &98\,927 & 722 & $98\,277+z_5$ & 1\,372\\
 & 33 & $^2P_{3/2}$ &         &  &  &  & 114\,830 & &114\,351&  & & \\
 & 34 & $^2P_{1/2}$ &         &  &  &  & 117\,607 & &117\,122 & & &\\
 & 35 & $^2D_{5/2}(1)$ &         &  &  &  & 128\,906 & &128\,281 & & &\\\vspace{2mm}
 & 36 & $^2D_{3/2}(1)$ &         &  &  &  & 130\,802 & &130\,180 & & &\\
10+ $4d^4$ & 0  & $^5D_0$ & $0+y$ & 0 & 0 & 0 & 0 & 0 & 0 & 0 & 0 & 0\\
 & 1  & $^5D_1$ & 3\,043    & 22 & 0 & 1 & 3\,043 & 0 & 3\,141 & -98 & 3\,035 & 8 \\
 & 2  & $^5D_2$ & $6\,590+x_5$  & 0 & 0 & 0 & 6\,590 &0  &6\,717 & -127 & 6\,545 & 45 \\
 & 3  & $^5D_3$ & 10\,073   & 49 & 84 & 1 & 10\,054 & 19 &10\,213 & -140 & 10\,005 & 68 \\
 & 4  & $^5D_4$ & 13\,300   & 24 & 70 & 3 & 13\,315 & -15 &13\,516 & -216 & 13\,280 & 20 \\
 & 5  & $^3P_{0}(2)$ & 26\,752   & 27 & 23 & 1 & 26\,750 & 2 &26\,490 & 262 & & \\
 & 6  & $^3H_4$ & 29\,584   & 15 & 75 & 1 & 29\,589 & -5 &29\,046 &538 & $29\,380+z_6$ & 204 \\
 & 7  & $^3G_3$ & $34\,918+x_5$ & 32 & 32 & 1 & 34\,899 & 19 &34\,573 & 345 & $34\,630+z_6$ & 288 \\
 & 8  & $^3H_5$ & 35\,147   & 24 & 73 & 1 & 35\,143 & 4 &34\,639 & 508 & $34\,814+z_7$ & 333 \\
 & 9  & $^3P_{1}(2)$ & $35\,425+x_5$ & 33 & 33 & 1 & 35\,429 & -4 &35\,438 & -13 & $35\,048+z_6$ & 377 \\
 & 10 & $^3F_{2}(2)$ & 36\,669   & 61 & 61 & 1 & 36\,666 & 3 &36\,613 & 56 & $36\,297+z_6$ & 372 \\
 & 11 & $^3H_6$ & 38\,232   & 10 & 70 & 1 & 38\,226 & 6 &37\,656 & 576 & $37\,890+z_7$ & 342 \\
 & 12 & $^3F_{4}(2)$ & 40\,490   & 13 & 70 & 2 & 40\,475 & 15 &40\,208 & 282 & $40\,130+z_7$ & 360 \\
 & 13 & $^3P_{2}(2)$ & 43\,351   & 44 & 95 & 1 & 43\,377 & -26 &43\,530 & -179 & $42\,898+z_6$ & 453 \\
 & 14 & $^3G_4$ & 43\,777   & 10 & 70 & 2 & 43\,765 & 12 &43\,539 & 238 & $43\,710+z_7$ & 67 \\
 & 15 & $^3F_{3}(2)$ & 45\,197   & 8  & 70 & 2 & 45\,196 & 1 &45\,146 & 51 & $44\,766+z_7$ & 431 \\
 & 16 & $^3D_3$ & 48\,279   & 35 & 80 & 2 & 48\,263 & 16 &48\,310 & -31 & $47\,850+z_7$ & 429 \\
 & 17 & $^3G_5$ & 48\,881   & 8 & 73 & 1 & 48\,893  & -12 &48\,663 & 218 & $48\,480+z_7$ & 401 \\
 & 18 & $^3D_2$ & 51\,801   & 30 & 60 & 1 & 51\,808 & -7 &51\,885 & -84 & & \\
 & 19 & $^3D_1$ & 52\,814   & 31 & 35 & 1 & 52\,806 & 8 &52\,945 & -131 & & \\
 & 20 & $^1I_6$ & 53\,860   & 14 & 74 & 1 & 53\,867 & -7 &53\,211 & 649 & $53\,475+z_7$ & 385 \\
 & 21 & $^1G_{4}(2)$ & 59\,693   & 6 & 70 & 4 & 59\,684  & 9 &59\,493 & 200 & & \\
 & 22 & $^1S_{0}(2)$ & $60\,890+x_5$ & 26 & 42 & 1 & 60\,870 & 20 &61\,041 & -151 & & \\
 & 23 & $^1D_{2}(2)$ & 64\,563   & 15 & 80 & 1 & 64\,549 & 14 &64\,542 & 21 & & \\
 & 24 & $^1F_3$ & 67\,957   & 16 & 80 & 1 & 67\,958 & -1 &67\,907 & 50 & $66\,757+z_8$ & 1\,200 \\
 & 25 & $^3P_{2}(1)$ &         &  &  &  & 75\,662 & &75\,823& & & \\
 & 26 & $^3F_{4}(1)$ & 78\,771   & 14 & 70 & 3 & 78\,777 & -6 &78\,854 & -83 & & \\
 & 27 & $^3F_{2}(1)$ & 80\,446   & 50 & 47 & 1 & 80\,445 & 1 &80\,484 & -38 & & \\
 & 28 & $^3F_{3}(1)$ & 81\,891   & 20 & 71 & 1 & 81\,881 & 10 &81\,982 & -91 & $80\,207+z_8$ & 1\,684 \\
 & 29 & $^3P_{1}(1)$ &         &  &  &  & 82\,941 & &83\,107& & & \\
 & 30 & $^3P_{0}(1)$ &         &  &  &  & 86\,664 & &86\,851 & & & \\
 & 31 & $^1G_{4}(1)$ &         &  &  &  & 89\,965 & &89\,627& & & \\
 & 32 & $^1D_{2}(1)$ &         &  &  &  & 112\,401 & &112\,544&  & & \\
 & \rule[-1.5mm]{0pt}{0mm}33 & $^1S_{0}(1)$ &         &  &  &  & 144\,549& &145\,002& & & \\
\end{longtable*}
\begin{table*}[t]
\caption{\label{tab:parameters}
Orthogonal energy parameters (all in cm$^{-1}$) obtained by fitting experimental energy levels (FIT) and ratios of FIT to \textsc{mcdf} code (FIT/MCDF) calculations. Experimental energy levels taken from: Ref.\,\cite{VanhetHof1993, azarov1994analysis} for Sn$^{6+}$, Ref.\,\cite{Azarov1993} for Sn$^{7+}$, this work for Sn$^{8+}$--Sn$^{10+}$ and Ref.\,\cite{Windberger2016} for Sn$^{11+}$, Sn$^{12+}$). Two-particle magnetic parameters ($A_c$-$A_0$) were fixed to \emph{ab initio} \textsc{mcdf} calculations, not fitted, and thus not listed here. Effective Coulomb-interaction operators $E_a'$, $E_b'$, and effective three-particle electrostatic operators $T$1 and $T$2 are fit parameters for the given number of $d$ electrons. For the fits parameters $E_b'$ in Sn$^{11+}$ and $E_a'$ in Sn$^{12+}$  were fixed (denoted by the superscript letter $f$) on extrapolated values (see main text). Fits were performed for the interacting 4$d^k$+4$d^{k-1}$5$s$+4$d^{k-2}$5$s^2$ configurations, $k$=8--2 for Sn$^{6+}$--Sn$^{12+}$ respectively. Energy parameters for the unknown 4$d^{k-1}$5$s$+4$d^{k-2}$5$s^2$ configurations (not listed) were fixed at \emph{ab initio} values for the average energies and spin-orbit interactions, but scaled by 0.85 for electrostatic and configuration interaction parameters. Average energy $E_{\mathrm{av}}$ is defined such that the ground level energy of the 4$d^k$+4$d^{k-1}$5$s$+4$d^{k-2}$5$s^2$ configurations is equal to zero. $\sigma$ is the root-mean-square of the fit uncertainty in cm$^{-1}$ for the calculated configuration. ``n/a'' indicates non-applicable parameters.}
\begin{tabular}{c r r r r r r r r}
\hline\hline
\multicolumn{1}{c}{Parameter} & \multicolumn{2}{c}{Sn$^{6+}$} & \multicolumn{2}{c}{Sn$^{7+}$} & \multicolumn{2}{c}{Sn$^{8+}$} & \multicolumn{2}{c}{Sn$^{9+}$} \\
 & \multicolumn{1}{c}{FIT} & \multicolumn{1}{c}{FIT/MCDF} & \multicolumn{1}{c}{FIT} & \multicolumn{1}{c}{FIT/MCDF} & \multicolumn{1}{c}{FIT} & \multicolumn{1}{c}{FIT/MCDF} & \multicolumn{1}{c}{FIT} & \multicolumn{1}{c}{FIT/MCDF} \\
\hline
$E_{\mathrm{av}}$   & 16\,279(4)   & 0.890       & 29\,983(4)   & 0.890      & 42\,688(4)   & 0.888   & 64\,568(10) & 0.872 \\
$O2$    & 9\,649(6)    & 0.856       & 9\,979(5)    & 0.860      & 10\,288(5)   & 0.863   & 10\,592(9)  & 0.867 \\
$O2'$   & 6\,100(6)    & 0.795       & 6\,326(4)    & 0.803     & 6\,526(7)    & 0.808   & 6\,702(11)  & 0.812  \\
$E_a'$   & 243(6)     &   &  247(3)     &      &  255(3)    &   &   256(6)       &       \\
$E_b'$   & 22(5)      &     &  34(3)     &      &  37(5)     &        &  66(10)         &         \\
$\zeta$(4$d$) & 3\,688(4)    & 1.024       & 3\,899(3)    & 1.021     & 4\,119(3)    & 1.020    & 4\,334(8)   & 1.018 \\
$T1$    & n/a  &  n/a  & -5.2(0.2)  &  & -5.6(0.1)  &         &    -5.9(0.3)  &           \\
$T2$    & n/a &     n/a     & n/a &     n/a      & 0.26(0.16)           &          &  -0.05(0.75)         &            \\
$\sigma$ & 10 & & 14 & & 16 & & 41 & \\
\hline\hline
\multicolumn{1}{c}{Parameter} & \multicolumn{2}{c}{Sn$^{10+}$} & \multicolumn{2}{c}{Sn$^{11+}$} & \multicolumn{2}{c}{Sn$^{12+}$} & \multicolumn{1}{c}{ }\\
 & \multicolumn{1}{c}{FIT} & \multicolumn{1}{c}{FIT/MCDF} & \multicolumn{1}{c}{FIT} & \multicolumn{1}{c}{FIT/MCDF} & \multicolumn{1}{c}{FIT} & \multicolumn{1}{c}{FIT/MCDF} & \multicolumn{1}{c}{}\\
\hline
$E_{\mathrm{av}}$   & 50\,180(3) & 0.91  & 37\,553(9)  & 0.911 & 24\,425(14) & 0.93  \\
$O2$    & 10\,878(3) & 0.87  & 11\,200(16) & 0.876 & 11\,480(43) & 0.88  \\
$O2'$   & 6\,896(6)  & 0.818 & 7\,007(23)  & 0.815 & 7\,102(43)  & 0.81  \\
$E_a'$  & 275(2)   &        & 248(15)    &      & 260$^f$    &       \\
$E_b'$  & 44(3)    &       & 50$^f$        &       &     n/a      &   n/a    \\
$\zeta$(4$d$) & 4563(3)  & 1.017 & 4783(6)   & 1.013 & 5038(11)  & 1.016 \\
$T1$     & -6.6(0.1)   &        & -9.72(1.0)     &      &      n/a     &  n/a \\
$T2$    & 0.37(0.16) &       &  n/a    &      n/a     &       n/a     &       n/a         \\
$\sigma$ & 14 & & 20 & & 32 &  \\
\hline\hline
\end{tabular}
\end{table*}
\newpage
\mbox{}
\newpage
\mbox{}
\newpage
\section*{References}
\bibliographystyle{apsrev4-1}
\bibliography{Optical_spectroscopy_of_complex_open_4d-shell_ions_Sn}

\begin{thebibliography}{63}%
\makeatletter
\providecommand \@ifxundefined [1]{%
 \@ifx{#1\undefined}
}%
\providecommand \@ifnum [1]{%
 \ifnum #1\expandafter \@firstoftwo
 \else \expandafter \@secondoftwo
 \fi
}%
\providecommand \@ifx [1]{%
 \ifx #1\expandafter \@firstoftwo
 \else \expandafter \@secondoftwo
 \fi
}%
\providecommand \natexlab [1]{#1}%
\providecommand \enquote  [1]{``#1''}%
\providecommand \bibnamefont  [1]{#1}%
\providecommand \bibfnamefont [1]{#1}%
\providecommand \citenamefont [1]{#1}%
\providecommand \href@noop [0]{\@secondoftwo}%
\providecommand \href [0]{\begingroup \@sanitize@url \@href}%
\providecommand \@href[1]{\@@startlink{#1}\@@href}%
\providecommand \@@href[1]{\endgroup#1\@@endlink}%
\providecommand \@sanitize@url [0]{\catcode `\\12\catcode `\$12\catcode
  `\&12\catcode `\#12\catcode `\^12\catcode `\_12\catcode `\%12\relax}%
\providecommand \@@startlink[1]{}%
\providecommand \@@endlink[0]{}%
\providecommand \url  [0]{\begingroup\@sanitize@url \@url }%
\providecommand \@url [1]{\endgroup\@href {#1}{\urlprefix }}%
\providecommand \urlprefix  [0]{URL }%
\providecommand \Eprint [0]{\href }%
\providecommand \doibase [0]{http://dx.doi.org/}%
\providecommand \selectlanguage [0]{\@gobble}%
\providecommand \bibinfo  [0]{\@secondoftwo}%
\providecommand \bibfield  [0]{\@secondoftwo}%
\providecommand \translation [1]{[#1]}%
\providecommand \BibitemOpen [0]{}%
\providecommand \bibitemStop [0]{}%
\providecommand \bibitemNoStop [0]{.\EOS\space}%
\providecommand \EOS [0]{\spacefactor3000\relax}%
\providecommand \BibitemShut  [1]{\csname bibitem#1\endcsname}%
\let\auto@bib@innerbib\@empty
\bibitem [{\citenamefont {Bauche}\ \emph {et~al.}(1988)\citenamefont {Bauche},
  \citenamefont {Bauche-Arnoult},\ and\ \citenamefont
  {Klapisch}}]{Bauche1988transition}%
  \BibitemOpen
  \bibfield  {author} {\bibinfo {author} {\bibfnamefont {J.}~\bibnamefont
  {Bauche}}, \bibinfo {author} {\bibfnamefont {C.}~\bibnamefont
  {Bauche-Arnoult}}, \ and\ \bibinfo {author} {\bibfnamefont {M.}~\bibnamefont
  {Klapisch}},\ }\href {\doibase 10.1016/S0065-2199(08)60107-4} {\bibfield
  {journal} {\bibinfo  {journal} {Adv. Atom. Mol. Phys.}\ }\textbf {\bibinfo
  {volume} {23}},\ \bibinfo {pages} {131} (\bibinfo {year} {1988})}\BibitemShut
  {NoStop}%
\bibitem [{\citenamefont {Fujioka}\ \emph {et~al.}(2008)\citenamefont
  {Fujioka}, \citenamefont {Shimomura}, \citenamefont {Shimada}, \citenamefont
  {Maeda}, \citenamefont {Sakaguchi}, \citenamefont {Nakai}, \citenamefont
  {Aota}, \citenamefont {Nishimura}, \citenamefont {Ozaki}, \citenamefont
  {Sunahara}, \citenamefont {Nishihara}, \citenamefont {Miyanaga},
  \citenamefont {Izawa},\ and\ \citenamefont {Mima}}]{Fujioka2008}%
  \BibitemOpen
  \bibfield  {author} {\bibinfo {author} {\bibfnamefont {S.}~\bibnamefont
  {Fujioka}}, \bibinfo {author} {\bibfnamefont {M.}~\bibnamefont {Shimomura}},
  \bibinfo {author} {\bibfnamefont {Y.}~\bibnamefont {Shimada}}, \bibinfo
  {author} {\bibfnamefont {S.}~\bibnamefont {Maeda}}, \bibinfo {author}
  {\bibfnamefont {H.}~\bibnamefont {Sakaguchi}}, \bibinfo {author}
  {\bibfnamefont {Y.}~\bibnamefont {Nakai}}, \bibinfo {author} {\bibfnamefont
  {T.}~\bibnamefont {Aota}}, \bibinfo {author} {\bibfnamefont {H.}~\bibnamefont
  {Nishimura}}, \bibinfo {author} {\bibfnamefont {N.}~\bibnamefont {Ozaki}},
  \bibinfo {author} {\bibfnamefont {A.}~\bibnamefont {Sunahara}}, \bibinfo
  {author} {\bibfnamefont {K.}~\bibnamefont {Nishihara}}, \bibinfo {author}
  {\bibfnamefont {N.}~\bibnamefont {Miyanaga}}, \bibinfo {author}
  {\bibfnamefont {Y.}~\bibnamefont {Izawa}}, \ and\ \bibinfo {author}
  {\bibfnamefont {K.}~\bibnamefont {Mima}},\ }\href {\doibase
  10.1063/1.2948874} {\bibfield  {journal} {\bibinfo  {journal} {Appl. Phys.
  Lett.}\ }\textbf {\bibinfo {volume} {92}},\ \bibinfo {pages} {241502}
  (\bibinfo {year} {2008})}\BibitemShut {NoStop}%
\bibitem [{\citenamefont {Benschop}\ \emph {et~al.}(2008)\citenamefont
  {Benschop}, \citenamefont {Banine}, \citenamefont {Lok},\ and\ \citenamefont
  {Loopstra}}]{Benschop2008}%
  \BibitemOpen
  \bibfield  {author} {\bibinfo {author} {\bibfnamefont {J.}~\bibnamefont
  {Benschop}}, \bibinfo {author} {\bibfnamefont {V.}~\bibnamefont {Banine}},
  \bibinfo {author} {\bibfnamefont {S.}~\bibnamefont {Lok}}, \ and\ \bibinfo
  {author} {\bibfnamefont {E.}~\bibnamefont {Loopstra}},\ }\href {\doibase
  10.1116/1.3010737} {\bibfield  {journal} {\bibinfo  {journal} {J. Vac. Sci.
  Technol. B}\ }\textbf {\bibinfo {volume} {26}},\ \bibinfo {pages} {2204}
  (\bibinfo {year} {2008})}\BibitemShut {NoStop}%
\bibitem [{\citenamefont {Mizoguchi}\ \emph {et~al.}(2010)\citenamefont
  {Mizoguchi}, \citenamefont {Abe}, \citenamefont {Watanabe}, \citenamefont
  {Ishihara}, \citenamefont {Ohta}, \citenamefont {Hori}, \citenamefont
  {Kurosu}, \citenamefont {Komori}, \citenamefont {Kakizaki}, \citenamefont
  {Sumitani}, \citenamefont {Wakabayashi}, \citenamefont {Nakarai},
  \citenamefont {Fujimoto},\ and\ \citenamefont {Endo}}]{Mizoguchi2010}%
  \BibitemOpen
  \bibfield  {author} {\bibinfo {author} {\bibfnamefont {H.}~\bibnamefont
  {Mizoguchi}}, \bibinfo {author} {\bibfnamefont {T.}~\bibnamefont {Abe}},
  \bibinfo {author} {\bibfnamefont {Y.}~\bibnamefont {Watanabe}}, \bibinfo
  {author} {\bibfnamefont {T.}~\bibnamefont {Ishihara}}, \bibinfo {author}
  {\bibfnamefont {T.}~\bibnamefont {Ohta}}, \bibinfo {author} {\bibfnamefont
  {T.}~\bibnamefont {Hori}}, \bibinfo {author} {\bibfnamefont {A.}~\bibnamefont
  {Kurosu}}, \bibinfo {author} {\bibfnamefont {H.}~\bibnamefont {Komori}},
  \bibinfo {author} {\bibfnamefont {K.}~\bibnamefont {Kakizaki}}, \bibinfo
  {author} {\bibfnamefont {A.}~\bibnamefont {Sumitani}}, \bibinfo {author}
  {\bibfnamefont {O.}~\bibnamefont {Wakabayashi}}, \bibinfo {author}
  {\bibfnamefont {H.}~\bibnamefont {Nakarai}}, \bibinfo {author} {\bibfnamefont
  {J.}~\bibnamefont {Fujimoto}}, \ and\ \bibinfo {author} {\bibfnamefont
  {A.}~\bibnamefont {Endo}},\ }\href {\doibase 10.1117/12.846271} {\bibfield
  {journal} {\bibinfo  {journal} {Proc. SPIE}\ }\textbf {\bibinfo {volume}
  {7636}},\ \bibinfo {pages} {763608} (\bibinfo {year} {2010})}\BibitemShut
  {NoStop}%
\bibitem [{\citenamefont {Banine}\ \emph {et~al.}(2011)\citenamefont {Banine},
  \citenamefont {Koshelev},\ and\ \citenamefont {Swinkels}}]{Banine2011}%
  \BibitemOpen
  \bibfield  {author} {\bibinfo {author} {\bibfnamefont {V.~Y.}\ \bibnamefont
  {Banine}}, \bibinfo {author} {\bibfnamefont {K.~N.}\ \bibnamefont
  {Koshelev}}, \ and\ \bibinfo {author} {\bibfnamefont {G.~H. P.~M.}\
  \bibnamefont {Swinkels}},\ }\href {\doibase 10.1088/0022-3727/44/25/253001}
  {\bibfield  {journal} {\bibinfo  {journal} {J. Phys. D: Appl. Phys.}\
  }\textbf {\bibinfo {volume} {44}},\ \bibinfo {pages} {253001} (\bibinfo
  {year} {2011})}\BibitemShut {NoStop}%
\bibitem [{\citenamefont {Azarov}\ and\ \citenamefont
  {Joshi}(1993)}]{Azarov1993}%
  \BibitemOpen
  \bibfield  {author} {\bibinfo {author} {\bibfnamefont {V.~I.}\ \bibnamefont
  {Azarov}}\ and\ \bibinfo {author} {\bibfnamefont {Y.~N.}\ \bibnamefont
  {Joshi}},\ }\href {\doibase 10.1088/0953-4075/26/20/011} {\bibfield
  {journal} {\bibinfo  {journal} {J. Phys. B}\ }\textbf {\bibinfo {volume}
  {26}},\ \bibinfo {pages} {3495} (\bibinfo {year} {1993})}\BibitemShut
  {NoStop}%
\bibitem [{\citenamefont {Tolstikhina}\ \emph {et~al.}(2006)\citenamefont
  {Tolstikhina}, \citenamefont {Churilov}, \citenamefont {Ryabtsev},\ and\
  \citenamefont {Koshelev}}]{tolstikhina2006ATOMICDATA}%
  \BibitemOpen
  \bibfield  {author} {\bibinfo {author} {\bibfnamefont {I.~Y.}\ \bibnamefont
  {Tolstikhina}}, \bibinfo {author} {\bibfnamefont {S.~S.}\ \bibnamefont
  {Churilov}}, \bibinfo {author} {\bibfnamefont {A.~N.}\ \bibnamefont
  {Ryabtsev}}, \ and\ \bibinfo {author} {\bibfnamefont {K.~N.}\ \bibnamefont
  {Koshelev}},\ }in\ \href@noop {} {\emph {\bibinfo {booktitle} {EUV sources
  for lithography}}},\ Vol.\ \bibinfo {volume} {149},\ \bibinfo {editor}
  {edited by\ \bibinfo {editor} {\bibfnamefont {V.}~\bibnamefont {Bakshi}}}\
  (\bibinfo  {publisher} {SPIE Press},\ \bibinfo {year} {2006})\ p.\ \bibinfo
  {pages} {113}\BibitemShut {NoStop}%
\bibitem [{\citenamefont {Svendsen}\ and\ \citenamefont
  {O'Sullivan}(1994)}]{Svendsen1994}%
  \BibitemOpen
  \bibfield  {author} {\bibinfo {author} {\bibfnamefont {W.}~\bibnamefont
  {Svendsen}}\ and\ \bibinfo {author} {\bibfnamefont {G.}~\bibnamefont
  {O'Sullivan}},\ }\href {\doibase 10.1103/PhysRevA.50.3710} {\bibfield
  {journal} {\bibinfo  {journal} {Phys. Rev. A}\ }\textbf {\bibinfo {volume}
  {50}},\ \bibinfo {pages} {3710} (\bibinfo {year} {1994})}\BibitemShut
  {NoStop}%
\bibitem [{\citenamefont {Sugar}\ \emph {et~al.}(1991)\citenamefont {Sugar},
  \citenamefont {Rowan},\ and\ \citenamefont {Kaufman}}]{sugar1991resonance}%
  \BibitemOpen
  \bibfield  {author} {\bibinfo {author} {\bibfnamefont {J.}~\bibnamefont
  {Sugar}}, \bibinfo {author} {\bibfnamefont {W.~L.}\ \bibnamefont {Rowan}}, \
  and\ \bibinfo {author} {\bibfnamefont {V.}~\bibnamefont {Kaufman}},\ }\href
  {\doibase 10.1364/JOSAB.8.002026} {\bibfield  {journal} {\bibinfo  {journal}
  {J. Opt. Soc. Am. B}\ }\textbf {\bibinfo {volume} {8}},\ \bibinfo {pages}
  {2026} (\bibinfo {year} {1991})}\BibitemShut {NoStop}%
\bibitem [{\citenamefont {Sugar}\ \emph {et~al.}(1992)\citenamefont {Sugar},
  \citenamefont {Rowan},\ and\ \citenamefont {Kaufman}}]{sugar1992rb}%
  \BibitemOpen
  \bibfield  {author} {\bibinfo {author} {\bibfnamefont {J.}~\bibnamefont
  {Sugar}}, \bibinfo {author} {\bibfnamefont {W.~L.}\ \bibnamefont {Rowan}}, \
  and\ \bibinfo {author} {\bibfnamefont {V.}~\bibnamefont {Kaufman}},\ }\href
  {\doibase 10.1364/JOSAB.9.001959} {\bibfield  {journal} {\bibinfo  {journal}
  {J. Opt. Soc. Am. B}\ }\textbf {\bibinfo {volume} {9}},\ \bibinfo {pages}
  {1959} (\bibinfo {year} {1992})}\BibitemShut {NoStop}%
\bibitem [{\citenamefont {Ohashi}\ \emph {et~al.}(2009)\citenamefont {Ohashi},
  \citenamefont {Suda}, \citenamefont {Tanuma}, \citenamefont {Fujioka},
  \citenamefont {Nishimura}, \citenamefont {Nishihara}, \citenamefont {Kai},
  \citenamefont {Sasaki}, \citenamefont {Sakaue}, \citenamefont {Nakamura}
  \emph {et~al.}}]{ohashi2009complementary}%
  \BibitemOpen
  \bibfield  {author} {\bibinfo {author} {\bibfnamefont {H.}~\bibnamefont
  {Ohashi}}, \bibinfo {author} {\bibfnamefont {S.}~\bibnamefont {Suda}},
  \bibinfo {author} {\bibfnamefont {H.}~\bibnamefont {Tanuma}}, \bibinfo
  {author} {\bibfnamefont {S.}~\bibnamefont {Fujioka}}, \bibinfo {author}
  {\bibfnamefont {H.}~\bibnamefont {Nishimura}}, \bibinfo {author}
  {\bibfnamefont {K.}~\bibnamefont {Nishihara}}, \bibinfo {author}
  {\bibfnamefont {T.}~\bibnamefont {Kai}}, \bibinfo {author} {\bibfnamefont
  {A.}~\bibnamefont {Sasaki}}, \bibinfo {author} {\bibfnamefont {H.~A.}\
  \bibnamefont {Sakaue}}, \bibinfo {author} {\bibfnamefont {N.}~\bibnamefont
  {Nakamura}},  \emph {et~al.},\ }in\ \href {\doibase
  10.1088/1742-6596/163/1/012071} {\emph {\bibinfo {booktitle} {J. Phys.: Conf.
  Ser.}}},\ Vol.\ \bibinfo {volume} {163}\ (\bibinfo {organization} {IOP
  Publishing},\ \bibinfo {year} {2009})\ p.\ \bibinfo {pages}
  {012071}\BibitemShut {NoStop}%
\bibitem [{\citenamefont {D'Arcy}\ \emph
  {et~al.}(2009{\natexlab{a}})\citenamefont {D'Arcy}, \citenamefont {Ohashi},
  \citenamefont {Suda}, \citenamefont {Tanuma}, \citenamefont {Fujioka},
  \citenamefont {Nishimura}, \citenamefont {Nishihara}, \citenamefont {Suzuki},
  \citenamefont {Kato}, \citenamefont {Koike}, \citenamefont {White},\ and\
  \citenamefont {O'Sullivan}}]{DArcy2009transitions}%
  \BibitemOpen
  \bibfield  {author} {\bibinfo {author} {\bibfnamefont {R.}~\bibnamefont
  {D'Arcy}}, \bibinfo {author} {\bibfnamefont {H.}~\bibnamefont {Ohashi}},
  \bibinfo {author} {\bibfnamefont {S.}~\bibnamefont {Suda}}, \bibinfo {author}
  {\bibfnamefont {H.}~\bibnamefont {Tanuma}}, \bibinfo {author} {\bibfnamefont
  {S.}~\bibnamefont {Fujioka}}, \bibinfo {author} {\bibfnamefont
  {H.}~\bibnamefont {Nishimura}}, \bibinfo {author} {\bibfnamefont
  {K.}~\bibnamefont {Nishihara}}, \bibinfo {author} {\bibfnamefont
  {C.}~\bibnamefont {Suzuki}}, \bibinfo {author} {\bibfnamefont
  {T.}~\bibnamefont {Kato}}, \bibinfo {author} {\bibfnamefont {F.}~\bibnamefont
  {Koike}}, \bibinfo {author} {\bibfnamefont {J.}~\bibnamefont {White}}, \ and\
  \bibinfo {author} {\bibfnamefont {G.}~\bibnamefont {O'Sullivan}},\ }\href
  {\doibase 10.1103/PhysRevA.79.042509} {\bibfield  {journal} {\bibinfo
  {journal} {Phys. Rev. A}\ }\textbf {\bibinfo {volume} {79}},\ \bibinfo
  {pages} {042509} (\bibinfo {year} {2009}{\natexlab{a}})}\BibitemShut
  {NoStop}%
\bibitem [{\citenamefont {D'Arcy}\ \emph
  {et~al.}(2009{\natexlab{b}})\citenamefont {D'Arcy}, \citenamefont {Ohashi},
  \citenamefont {Suda}, \citenamefont {Tanuma}, \citenamefont {Fujioka},
  \citenamefont {Nishimura}, \citenamefont {Nishihara}, \citenamefont {Suzuki},
  \citenamefont {Kato}, \citenamefont {Koike}, \citenamefont {O'Connor},\ and\
  \citenamefont {O'Sullivan}}]{DArcy2009identification}%
  \BibitemOpen
  \bibfield  {author} {\bibinfo {author} {\bibfnamefont {R.}~\bibnamefont
  {D'Arcy}}, \bibinfo {author} {\bibfnamefont {H.}~\bibnamefont {Ohashi}},
  \bibinfo {author} {\bibfnamefont {S.}~\bibnamefont {Suda}}, \bibinfo {author}
  {\bibfnamefont {H.}~\bibnamefont {Tanuma}}, \bibinfo {author} {\bibfnamefont
  {S.}~\bibnamefont {Fujioka}}, \bibinfo {author} {\bibfnamefont
  {H.}~\bibnamefont {Nishimura}}, \bibinfo {author} {\bibfnamefont
  {K.}~\bibnamefont {Nishihara}}, \bibinfo {author} {\bibfnamefont
  {C.}~\bibnamefont {Suzuki}}, \bibinfo {author} {\bibfnamefont
  {T.}~\bibnamefont {Kato}}, \bibinfo {author} {\bibfnamefont {F.}~\bibnamefont
  {Koike}}, \bibinfo {author} {\bibfnamefont {A.}~\bibnamefont {O'Connor}}, \
  and\ \bibinfo {author} {\bibfnamefont {G.}~\bibnamefont {O'Sullivan}},\
  }\href {\doibase 10.1088/0953-4075/42/16/165207} {\bibfield  {journal}
  {\bibinfo  {journal} {J. Phys. B}\ }\textbf {\bibinfo {volume} {42}},\
  \bibinfo {pages} {165207} (\bibinfo {year} {2009}{\natexlab{b}})}\BibitemShut
  {NoStop}%
\bibitem [{\citenamefont {Ohashi}\ \emph {et~al.}(2010)\citenamefont {Ohashi},
  \citenamefont {Suda}, \citenamefont {Tanuma}, \citenamefont {Fujioka},
  \citenamefont {Nishimura}, \citenamefont {Sasaki},\ and\ \citenamefont
  {Nishihara}}]{ohashi2010euv}%
  \BibitemOpen
  \bibfield  {author} {\bibinfo {author} {\bibfnamefont {H.}~\bibnamefont
  {Ohashi}}, \bibinfo {author} {\bibfnamefont {S.}~\bibnamefont {Suda}},
  \bibinfo {author} {\bibfnamefont {H.}~\bibnamefont {Tanuma}}, \bibinfo
  {author} {\bibfnamefont {S.}~\bibnamefont {Fujioka}}, \bibinfo {author}
  {\bibfnamefont {H.}~\bibnamefont {Nishimura}}, \bibinfo {author}
  {\bibfnamefont {A.}~\bibnamefont {Sasaki}}, \ and\ \bibinfo {author}
  {\bibfnamefont {K.}~\bibnamefont {Nishihara}},\ }\href {\doibase
  10.1088/0953-4075/43/6/065204} {\bibfield  {journal} {\bibinfo  {journal} {J.
  Phys. B}\ }\textbf {\bibinfo {volume} {43}},\ \bibinfo {pages} {065204}
  (\bibinfo {year} {2010})}\BibitemShut {NoStop}%
\bibitem [{\citenamefont {Yatsurugi}\ \emph {et~al.}(2011)\citenamefont
  {Yatsurugi}, \citenamefont {Watanabe}, \citenamefont {Ohashi}, \citenamefont
  {Sakaue},\ and\ \citenamefont {Nakamura}}]{yatsurugi2011euv}%
  \BibitemOpen
  \bibfield  {author} {\bibinfo {author} {\bibfnamefont {J.}~\bibnamefont
  {Yatsurugi}}, \bibinfo {author} {\bibfnamefont {E.}~\bibnamefont {Watanabe}},
  \bibinfo {author} {\bibfnamefont {H.}~\bibnamefont {Ohashi}}, \bibinfo
  {author} {\bibfnamefont {H.~A.}\ \bibnamefont {Sakaue}}, \ and\ \bibinfo
  {author} {\bibfnamefont {N.}~\bibnamefont {Nakamura}},\ }\href {\doibase
  10.1088/0031-8949/2011/T144/014031} {\bibfield  {journal} {\bibinfo
  {journal} {Phys. Scr.}\ }\textbf {\bibinfo {volume} {2011}},\ \bibinfo
  {pages} {014031} (\bibinfo {year} {2011})}\BibitemShut {NoStop}%
\bibitem [{\citenamefont {Windberger}\ \emph {et~al.}(2016)\citenamefont
  {Windberger}, \citenamefont {Torretti}, \citenamefont {Borschevsky},
  \citenamefont {Ryabtsev}, \citenamefont {Dobrodey}, \citenamefont {Bekker},
  \citenamefont {Eliav}, \citenamefont {Kaldor}, \citenamefont {Ubachs},
  \citenamefont {Hoekstra}, \citenamefont {Crespo L\'opez-Urrutia},\ and\
  \citenamefont {Versolato}}]{Windberger2016}%
  \BibitemOpen
  \bibfield  {author} {\bibinfo {author} {\bibfnamefont {A.}~\bibnamefont
  {Windberger}}, \bibinfo {author} {\bibfnamefont {F.}~\bibnamefont
  {Torretti}}, \bibinfo {author} {\bibfnamefont {A.}~\bibnamefont
  {Borschevsky}}, \bibinfo {author} {\bibfnamefont {A.}~\bibnamefont
  {Ryabtsev}}, \bibinfo {author} {\bibfnamefont {S.}~\bibnamefont {Dobrodey}},
  \bibinfo {author} {\bibfnamefont {H.}~\bibnamefont {Bekker}}, \bibinfo
  {author} {\bibfnamefont {E.}~\bibnamefont {Eliav}}, \bibinfo {author}
  {\bibfnamefont {U.}~\bibnamefont {Kaldor}}, \bibinfo {author} {\bibfnamefont
  {W.}~\bibnamefont {Ubachs}}, \bibinfo {author} {\bibfnamefont
  {R.}~\bibnamefont {Hoekstra}}, \bibinfo {author} {\bibfnamefont {J.~R.}\
  \bibnamefont {Crespo L\'opez-Urrutia}}, \ and\ \bibinfo {author}
  {\bibfnamefont {O.~O.}\ \bibnamefont {Versolato}},\ }\href {\doibase
  10.1103/PhysRevA.94.012506} {\bibfield  {journal} {\bibinfo  {journal} {Phys.
  Rev. A}\ }\textbf {\bibinfo {volume} {94}},\ \bibinfo {pages} {012506}
  (\bibinfo {year} {2016})}\BibitemShut {NoStop}%
\bibitem [{\citenamefont {Churilov}\ and\ \citenamefont
  {Ryabtsev}(2006{\natexlab{a}})}]{Churilov2006SnVIII}%
  \BibitemOpen
  \bibfield  {author} {\bibinfo {author} {\bibfnamefont {S.~S.}\ \bibnamefont
  {Churilov}}\ and\ \bibinfo {author} {\bibfnamefont {A.~N.}\ \bibnamefont
  {Ryabtsev}},\ }\href {\doibase 10.1134/S0030400X06050043} {\bibfield
  {journal} {\bibinfo  {journal} {Opt. Spectrosc.}\ }\textbf {\bibinfo {volume}
  {100}},\ \bibinfo {pages} {660} (\bibinfo {year}
  {2006}{\natexlab{a}})}\BibitemShut {NoStop}%
\bibitem [{\citenamefont {Churilov}\ and\ \citenamefont
  {Ryabtsev}(2006{\natexlab{b}})}]{Churilov2006SnIX--SnXII}%
  \BibitemOpen
  \bibfield  {author} {\bibinfo {author} {\bibfnamefont {S.~S.}\ \bibnamefont
  {Churilov}}\ and\ \bibinfo {author} {\bibfnamefont {A.~N.}\ \bibnamefont
  {Ryabtsev}},\ }\href {\doibase 10.1088/0031-8949/73/6/014} {\bibfield
  {journal} {\bibinfo  {journal} {Phys. Scr.}\ }\textbf {\bibinfo {volume}
  {73}},\ \bibinfo {pages} {614} (\bibinfo {year}
  {2006}{\natexlab{b}})}\BibitemShut {NoStop}%
\bibitem [{\citenamefont {Churilov}\ and\ \citenamefont
  {Ryabtsev}(2006{\natexlab{c}})}]{Churilov2006SnXIII--XV}%
  \BibitemOpen
  \bibfield  {author} {\bibinfo {author} {\bibfnamefont {S.~S.}\ \bibnamefont
  {Churilov}}\ and\ \bibinfo {author} {\bibfnamefont {A.~N.}\ \bibnamefont
  {Ryabtsev}},\ }\href {\doibase 10.1134/S0030400X06080017} {\bibfield
  {journal} {\bibinfo  {journal} {Opt. Spectrosc.}\ }\textbf {\bibinfo {volume}
  {101}},\ \bibinfo {pages} {169} (\bibinfo {year}
  {2006}{\natexlab{c}})}\BibitemShut {NoStop}%
\bibitem [{\citenamefont {Ryabtsev}\ \emph {et~al.}(2008)\citenamefont
  {Ryabtsev}, \citenamefont {Kononov},\ and\ \citenamefont
  {Churilov}}]{ryabtsev2008SnXIV}%
  \BibitemOpen
  \bibfield  {author} {\bibinfo {author} {\bibfnamefont {A.~N.}\ \bibnamefont
  {Ryabtsev}}, \bibinfo {author} {\bibfnamefont {{\'E}.~Y.}\ \bibnamefont
  {Kononov}}, \ and\ \bibinfo {author} {\bibfnamefont {S.~S.}\ \bibnamefont
  {Churilov}},\ }\href {\doibase 10.1134/S0030400X08120060} {\bibfield
  {journal} {\bibinfo  {journal} {Opt. Spectrosc.}\ }\textbf {\bibinfo {volume}
  {105}},\ \bibinfo {pages} {844} (\bibinfo {year} {2008})}\BibitemShut
  {NoStop}%
\bibitem [{\citenamefont {Flambaum}\ \emph {et~al.}(1994)\citenamefont
  {Flambaum}, \citenamefont {Gribakina}, \citenamefont {Gribakin},\ and\
  \citenamefont {Kozlov}}]{flambaum1994structure}%
  \BibitemOpen
  \bibfield  {author} {\bibinfo {author} {\bibfnamefont {V.}~\bibnamefont
  {Flambaum}}, \bibinfo {author} {\bibfnamefont {A.}~\bibnamefont {Gribakina}},
  \bibinfo {author} {\bibfnamefont {G.}~\bibnamefont {Gribakin}}, \ and\
  \bibinfo {author} {\bibfnamefont {M.}~\bibnamefont {Kozlov}},\ }\href
  {\doibase 10.1103/PhysRevA.50.267} {\bibfield  {journal} {\bibinfo  {journal}
  {Phys. Rev. A}\ }\textbf {\bibinfo {volume} {50}},\ \bibinfo {pages} {267}
  (\bibinfo {year} {1994})}\BibitemShut {NoStop}%
\bibitem [{\citenamefont {Dzuba}\ \emph {et~al.}(2012)\citenamefont {Dzuba},
  \citenamefont {Flambaum}, \citenamefont {Gribakin},\ and\ \citenamefont
  {Harabati}}]{dzuba2012chaos}%
  \BibitemOpen
  \bibfield  {author} {\bibinfo {author} {\bibfnamefont {V.}~\bibnamefont
  {Dzuba}}, \bibinfo {author} {\bibfnamefont {V.}~\bibnamefont {Flambaum}},
  \bibinfo {author} {\bibfnamefont {G.}~\bibnamefont {Gribakin}}, \ and\
  \bibinfo {author} {\bibfnamefont {C.}~\bibnamefont {Harabati}},\ }\href
  {\doibase 10.1103/PhysRevA.86.022714} {\bibfield  {journal} {\bibinfo
  {journal} {Phys. Rev. A}\ }\textbf {\bibinfo {volume} {86}},\ \bibinfo
  {pages} {022714} (\bibinfo {year} {2012})}\BibitemShut {NoStop}%
\bibitem [{\citenamefont {Berengut}\ \emph {et~al.}(2015)\citenamefont
  {Berengut}, \citenamefont {Harabati}, \citenamefont {Dzuba}, \citenamefont
  {Flambaum},\ and\ \citenamefont {Gribakin}}]{berengut2015level}%
  \BibitemOpen
  \bibfield  {author} {\bibinfo {author} {\bibfnamefont {J.}~\bibnamefont
  {Berengut}}, \bibinfo {author} {\bibfnamefont {C.}~\bibnamefont {Harabati}},
  \bibinfo {author} {\bibfnamefont {V.}~\bibnamefont {Dzuba}}, \bibinfo
  {author} {\bibfnamefont {V.}~\bibnamefont {Flambaum}}, \ and\ \bibinfo
  {author} {\bibfnamefont {G.}~\bibnamefont {Gribakin}},\ }\href {\doibase
  10.1103/PhysRevA.92.062717} {\bibfield  {journal} {\bibinfo  {journal} {Phys.
  Rev. A}\ }\textbf {\bibinfo {volume} {92}},\ \bibinfo {pages} {062717}
  (\bibinfo {year} {2015})}\BibitemShut {NoStop}%
\bibitem [{\citenamefont {Harabati}\ \emph {et~al.}(2016)\citenamefont
  {Harabati}, \citenamefont {Berengut}, \citenamefont {Flambaum},\ and\
  \citenamefont {Dzuba}}]{harabati2016electron}%
  \BibitemOpen
  \bibfield  {author} {\bibinfo {author} {\bibfnamefont {C.}~\bibnamefont
  {Harabati}}, \bibinfo {author} {\bibfnamefont {J.}~\bibnamefont {Berengut}},
  \bibinfo {author} {\bibfnamefont {V.}~\bibnamefont {Flambaum}}, \ and\
  \bibinfo {author} {\bibfnamefont {V.}~\bibnamefont {Dzuba}},\ }\href
  {https://arxiv.org/abs/1608.07932} {\bibfield  {journal} {\bibinfo  {journal}
  {arXiv preprint arXiv:1608.07932}\ } (\bibinfo {year} {2016})}\BibitemShut
  {NoStop}%
\bibitem [{\citenamefont {Dzuba}\ \emph {et~al.}(1996)\citenamefont {Dzuba},
  \citenamefont {Flambaum},\ and\ \citenamefont {Kozlov}}]{dzuba1996}%
  \BibitemOpen
  \bibfield  {author} {\bibinfo {author} {\bibfnamefont {V.~A.}\ \bibnamefont
  {Dzuba}}, \bibinfo {author} {\bibfnamefont {V.~V.}\ \bibnamefont {Flambaum}},
  \ and\ \bibinfo {author} {\bibfnamefont {M.~G.}\ \bibnamefont {Kozlov}},\
  }\href {\doibase 10.1103/PhysRevA.54.3948} {\bibfield  {journal} {\bibinfo
  {journal} {Phys. Rev. A}\ }\textbf {\bibinfo {volume} {54}},\ \bibinfo
  {pages} {3948} (\bibinfo {year} {1996})}\BibitemShut {NoStop}%
\bibitem [{\citenamefont {Berengut}\ \emph {et~al.}(2006)\citenamefont
  {Berengut}, \citenamefont {Flambaum},\ and\ \citenamefont
  {Kozlov}}]{berengut06pra}%
  \BibitemOpen
  \bibfield  {author} {\bibinfo {author} {\bibfnamefont {J.~C.}\ \bibnamefont
  {Berengut}}, \bibinfo {author} {\bibfnamefont {V.~V.}\ \bibnamefont
  {Flambaum}}, \ and\ \bibinfo {author} {\bibfnamefont {M.~G.}\ \bibnamefont
  {Kozlov}},\ }\href {\doibase 10.1103/PhysRevA.73.012504} {\bibfield
  {journal} {\bibinfo  {journal} {Phys. Rev. A}\ }\textbf {\bibinfo {volume}
  {73}},\ \bibinfo {pages} {012504} (\bibinfo {year} {2006})}\BibitemShut
  {NoStop}%
\bibitem [{\citenamefont {Berengut}\ \emph {et~al.}(2013)\citenamefont
  {Berengut}, \citenamefont {Flambaum}, \citenamefont {Ong}, \citenamefont
  {Webb}, \citenamefont {Barrow}, \citenamefont {Barstow}, \citenamefont
  {Preval},\ and\ \citenamefont {Holberg}}]{berengut13prl}%
  \BibitemOpen
  \bibfield  {author} {\bibinfo {author} {\bibfnamefont {J.~C.}\ \bibnamefont
  {Berengut}}, \bibinfo {author} {\bibfnamefont {V.~V.}\ \bibnamefont
  {Flambaum}}, \bibinfo {author} {\bibfnamefont {A.}~\bibnamefont {Ong}},
  \bibinfo {author} {\bibfnamefont {J.~K.}\ \bibnamefont {Webb}}, \bibinfo
  {author} {\bibfnamefont {J.~D.}\ \bibnamefont {Barrow}}, \bibinfo {author}
  {\bibfnamefont {M.~A.}\ \bibnamefont {Barstow}}, \bibinfo {author}
  {\bibfnamefont {S.~P.}\ \bibnamefont {Preval}}, \ and\ \bibinfo {author}
  {\bibfnamefont {J.~B.}\ \bibnamefont {Holberg}},\ }\href {\doibase
  10.1103/PhysRevLett.111.010801} {\bibfield  {journal} {\bibinfo  {journal}
  {Phys. Rev. Lett.}\ }\textbf {\bibinfo {volume} {111}},\ \bibinfo {pages}
  {010801} (\bibinfo {year} {2013})}\BibitemShut {NoStop}%
\bibitem [{\citenamefont {Ong}\ \emph {et~al.}(2013)\citenamefont {Ong},
  \citenamefont {Berengut},\ and\ \citenamefont {Flambaum}}]{ong13pra}%
  \BibitemOpen
  \bibfield  {author} {\bibinfo {author} {\bibfnamefont {A.}~\bibnamefont
  {Ong}}, \bibinfo {author} {\bibfnamefont {J.~C.}\ \bibnamefont {Berengut}}, \
  and\ \bibinfo {author} {\bibfnamefont {V.~V.}\ \bibnamefont {Flambaum}},\
  }\href {\doibase 10.1103/PhysRevA.88.052517} {\bibfield  {journal} {\bibinfo
  {journal} {Phys. Rev. A}\ }\textbf {\bibinfo {volume} {88}},\ \bibinfo
  {pages} {052517} (\bibinfo {year} {2013})}\BibitemShut {NoStop}%
\bibitem [{\citenamefont {Savukov}(2015)}]{savukov_CI_2015}%
  \BibitemOpen
  \bibfield  {author} {\bibinfo {author} {\bibfnamefont {I.~M.}\ \bibnamefont
  {Savukov}},\ }\href {\doibase 10.1103/PhysRevA.91.022514} {\bibfield
  {journal} {\bibinfo  {journal} {Phys. Rev. A}\ }\textbf {\bibinfo {volume}
  {91}},\ \bibinfo {pages} {022514} (\bibinfo {year} {2015})}\BibitemShut
  {NoStop}%
\bibitem [{\citenamefont {Berengut}(2011)}]{berengut_five_2011}%
  \BibitemOpen
  \bibfield  {author} {\bibinfo {author} {\bibfnamefont {J.~C.}\ \bibnamefont
  {Berengut}},\ }\href {\doibase 10.1103/PhysRevA.84.052520} {\bibfield
  {journal} {\bibinfo  {journal} {Phys. Rev. A}\ }\textbf {\bibinfo {volume}
  {84}},\ \bibinfo {pages} {052520} (\bibinfo {year} {2011})}\BibitemShut
  {NoStop}%
\bibitem [{\citenamefont {Berengut}(2016)}]{berengut_CIMBPT_2016}%
  \BibitemOpen
  \bibfield  {author} {\bibinfo {author} {\bibfnamefont {J.~C.}\ \bibnamefont
  {Berengut}},\ }\href {\doibase 10.1103/PhysRevA.94.012502} {\bibfield
  {journal} {\bibinfo  {journal} {Phys. Rev. A}\ }\textbf {\bibinfo {volume}
  {94}},\ \bibinfo {pages} {012502} (\bibinfo {year} {2016})}\BibitemShut
  {NoStop}%
\bibitem [{\citenamefont {Epp}\ \emph {et~al.}(2010)\citenamefont {Epp},
  \citenamefont {{J. R. Crespo L\'{o}pez-Urrutia}}, \citenamefont {Simon},
  \citenamefont {Baumann}, \citenamefont {Brenner}, \citenamefont {Ginzel},
  \citenamefont {Guerassimova}, \citenamefont {M{\"a}ckel}, \citenamefont
  {Mokler}, \citenamefont {Schmitt} \emph {et~al.}}]{epp2010x}%
  \BibitemOpen
  \bibfield  {author} {\bibinfo {author} {\bibfnamefont {S.~W.}\ \bibnamefont
  {Epp}}, \bibinfo {author} {\bibnamefont {{J. R. Crespo L\'{o}pez-Urrutia}}},
  \bibinfo {author} {\bibfnamefont {M.~C.}\ \bibnamefont {Simon}}, \bibinfo
  {author} {\bibfnamefont {T.}~\bibnamefont {Baumann}}, \bibinfo {author}
  {\bibfnamefont {G.}~\bibnamefont {Brenner}}, \bibinfo {author} {\bibfnamefont
  {R.}~\bibnamefont {Ginzel}}, \bibinfo {author} {\bibfnamefont
  {N.}~\bibnamefont {Guerassimova}}, \bibinfo {author} {\bibfnamefont
  {V.}~\bibnamefont {M{\"a}ckel}}, \bibinfo {author} {\bibfnamefont {P.~H.}\
  \bibnamefont {Mokler}}, \bibinfo {author} {\bibfnamefont {B.~L.}\
  \bibnamefont {Schmitt}},  \emph {et~al.},\ }\href {\doibase
  10.1088/0953-4075/43/19/194008} {\bibfield  {journal} {\bibinfo  {journal}
  {J. Phys. B}\ }\textbf {\bibinfo {volume} {43}},\ \bibinfo {pages} {194008}
  (\bibinfo {year} {2010})}\BibitemShut {NoStop}%
\bibitem [{\citenamefont {Uylings}\ \emph {et~al.}(1993)\citenamefont
  {Uylings}, \citenamefont {Raassen},\ and\ \citenamefont
  {Wyart}}]{Uylings1993}%
  \BibitemOpen
  \bibfield  {author} {\bibinfo {author} {\bibfnamefont {P.}~\bibnamefont
  {Uylings}}, \bibinfo {author} {\bibfnamefont {A.}~\bibnamefont {Raassen}}, \
  and\ \bibinfo {author} {\bibfnamefont {J.-F.}\ \bibnamefont {Wyart}},\ }\href
  {\doibase 10.1088/0953-4075/26/24/004} {\bibfield  {journal} {\bibinfo
  {journal} {J. Phys. B}\ }\textbf {\bibinfo {volume} {26}},\ \bibinfo {pages}
  {4683} (\bibinfo {year} {1993})}\BibitemShut {NoStop}%
\bibitem [{\citenamefont {Hansen}\ \emph
  {et~al.}(1988{\natexlab{a}})\citenamefont {Hansen}, \citenamefont {Uylings},\
  and\ \citenamefont {Raassen}}]{Hansen1988a}%
  \BibitemOpen
  \bibfield  {author} {\bibinfo {author} {\bibfnamefont {J.}~\bibnamefont
  {Hansen}}, \bibinfo {author} {\bibfnamefont {P.}~\bibnamefont {Uylings}}, \
  and\ \bibinfo {author} {\bibfnamefont {A.}~\bibnamefont {Raassen}},\ }\href
  {\doibase 10.1088/0031-8949/37/5/003} {\bibfield  {journal} {\bibinfo
  {journal} {Phys. Scr.}\ }\textbf {\bibinfo {volume} {37}},\ \bibinfo {pages}
  {664} (\bibinfo {year} {1988}{\natexlab{a}})}\BibitemShut {NoStop}%
\bibitem [{\citenamefont {Cowan}(1981)}]{cowan1981}%
  \BibitemOpen
  \bibfield  {author} {\bibinfo {author} {\bibfnamefont {R.~D.}\ \bibnamefont
  {Cowan}},\ }\href@noop {} {\emph {\bibinfo {title} {{The Theory of Atomic
  Structure and Spectra}}}}\ (\bibinfo  {publisher} {University of California
  Press},\ \bibinfo {year} {1981})\BibitemShut {NoStop}%
\bibitem [{\citenamefont {Bekker}\ \emph {et~al.}(2015)\citenamefont {Bekker},
  \citenamefont {Versolato}, \citenamefont {Windberger}, \citenamefont
  {Oreshkina}, \citenamefont {Schupp}, \citenamefont {Baumann}, \citenamefont
  {Harman}, \citenamefont {Keitel}, \citenamefont {Schmidt}, \citenamefont
  {Ullrich},\ and\ \citenamefont {{J. R. Crespo
  L\'{o}pez-Urrutia}}}]{Bekker2015a}%
  \BibitemOpen
  \bibfield  {author} {\bibinfo {author} {\bibfnamefont {H.}~\bibnamefont
  {Bekker}}, \bibinfo {author} {\bibfnamefont {O.~O.}\ \bibnamefont
  {Versolato}}, \bibinfo {author} {\bibfnamefont {A.}~\bibnamefont
  {Windberger}}, \bibinfo {author} {\bibfnamefont {N.~S.}\ \bibnamefont
  {Oreshkina}}, \bibinfo {author} {\bibfnamefont {R.}~\bibnamefont {Schupp}},
  \bibinfo {author} {\bibfnamefont {T.~M.}\ \bibnamefont {Baumann}}, \bibinfo
  {author} {\bibfnamefont {Z.}~\bibnamefont {Harman}}, \bibinfo {author}
  {\bibfnamefont {C.~H.}\ \bibnamefont {Keitel}}, \bibinfo {author}
  {\bibfnamefont {P.~O.}\ \bibnamefont {Schmidt}}, \bibinfo {author}
  {\bibfnamefont {J.}~\bibnamefont {Ullrich}}, \ and\ \bibinfo {author}
  {\bibnamefont {{J. R. Crespo L\'{o}pez-Urrutia}}},\ }\href {\doibase
  10.1088/0953-4075/48/14/144018} {\bibfield  {journal} {\bibinfo  {journal}
  {J. Phys. B}\ }\textbf {\bibinfo {volume} {48}},\ \bibinfo {pages} {144018}
  (\bibinfo {year} {2015})}\BibitemShut {NoStop}%
\bibitem [{\citenamefont {Harada}\ and\ \citenamefont
  {Kita}(1980)}]{Harada1980}%
  \BibitemOpen
  \bibfield  {author} {\bibinfo {author} {\bibfnamefont {T.}~\bibnamefont
  {Harada}}\ and\ \bibinfo {author} {\bibfnamefont {T.}~\bibnamefont {Kita}},\
  }\href {\doibase 10.1364/AO.19.003987} {\bibfield  {journal} {\bibinfo
  {journal} {Appl. Opt.}\ }\textbf {\bibinfo {volume} {19}},\ \bibinfo {pages}
  {3987} (\bibinfo {year} {1980})}\BibitemShut {NoStop}%
\bibitem [{\citenamefont {Johnson}\ \emph {et~al.}(1988)\citenamefont
  {Johnson}, \citenamefont {Blundell},\ and\ \citenamefont
  {Sapirstein}}]{johnson88pra}%
  \BibitemOpen
  \bibfield  {author} {\bibinfo {author} {\bibfnamefont {W.~R.}\ \bibnamefont
  {Johnson}}, \bibinfo {author} {\bibfnamefont {S.~A.}\ \bibnamefont
  {Blundell}}, \ and\ \bibinfo {author} {\bibfnamefont {J.}~\bibnamefont
  {Sapirstein}},\ }\href {\doibase 10.1103/PhysRevA.37.307} {\bibfield
  {journal} {\bibinfo  {journal} {Phys. Rev. A}\ }\textbf {\bibinfo {volume}
  {37}},\ \bibinfo {pages} {307} (\bibinfo {year} {1988})}\BibitemShut
  {NoStop}%
\bibitem [{\citenamefont {Dzuba}\ and\ \citenamefont
  {Johnson}(1998)}]{dzuba1998}%
  \BibitemOpen
  \bibfield  {author} {\bibinfo {author} {\bibfnamefont {V.}~\bibnamefont
  {Dzuba}}\ and\ \bibinfo {author} {\bibfnamefont {W.}~\bibnamefont
  {Johnson}},\ }\href {\doibase 10.1103/PhysRevA.57.2459} {\bibfield  {journal}
  {\bibinfo  {journal} {Physical Review A}\ }\textbf {\bibinfo {volume} {57}},\
  \bibinfo {pages} {2459} (\bibinfo {year} {1998})}\BibitemShut {NoStop}%
\bibitem [{\citenamefont {Beloy}\ and\ \citenamefont
  {Derevianko}(2008)}]{beloy08cpc}%
  \BibitemOpen
  \bibfield  {author} {\bibinfo {author} {\bibfnamefont {K.}~\bibnamefont
  {Beloy}}\ and\ \bibinfo {author} {\bibfnamefont {A.}~\bibnamefont
  {Derevianko}},\ }\href {\doibase 10.1016/j.cpc.2008.03.004} {\bibfield
  {journal} {\bibinfo  {journal} {Comp. Phys. Comm.}\ }\textbf {\bibinfo
  {volume} {179}},\ \bibinfo {pages} {310} (\bibinfo {year}
  {2008})}\BibitemShut {NoStop}%
\bibitem [{\citenamefont {Kozlov}\ and\ \citenamefont
  {Porsev}(1999)}]{kozlov99os}%
  \BibitemOpen
  \bibfield  {author} {\bibinfo {author} {\bibfnamefont {M.~G.}\ \bibnamefont
  {Kozlov}}\ and\ \bibinfo {author} {\bibfnamefont {S.~G.}\ \bibnamefont
  {Porsev}},\ }\href@noop {} {\bibfield  {journal} {\bibinfo  {journal} {Optics
  and Spectroscopy}\ }\textbf {\bibinfo {volume} {87}},\ \bibinfo {pages} {352}
  (\bibinfo {year} {1999})}\BibitemShut {NoStop}%
\bibitem [{\citenamefont {Dzuba}(2005)}]{dzuba05pra1}%
  \BibitemOpen
  \bibfield  {author} {\bibinfo {author} {\bibfnamefont {V.~A.}\ \bibnamefont
  {Dzuba}},\ }\href {\doibase 10.1103/PhysRevA.71.032512} {\bibfield  {journal}
  {\bibinfo  {journal} {Phys. Rev. A}\ }\textbf {\bibinfo {volume} {71}},\
  \bibinfo {pages} {032512} (\bibinfo {year} {2005})}\BibitemShut {NoStop}%
\bibitem [{\citenamefont {Ginges}\ and\ \citenamefont
  {Berengut}(2016{\natexlab{a}})}]{ginges16jpb}%
  \BibitemOpen
  \bibfield  {author} {\bibinfo {author} {\bibfnamefont {J.~S.~M.}\
  \bibnamefont {Ginges}}\ and\ \bibinfo {author} {\bibfnamefont {J.~C.}\
  \bibnamefont {Berengut}},\ }\href {\doibase 10.1088/0953-4075/49/9/095001}
  {\bibfield  {journal} {\bibinfo  {journal} {J. Phys. B}\ }\textbf {\bibinfo
  {volume} {49}},\ \bibinfo {pages} {095001} (\bibinfo {year}
  {2016}{\natexlab{a}})}\BibitemShut {NoStop}%
\bibitem [{\citenamefont {Ginges}\ and\ \citenamefont
  {Berengut}(2016{\natexlab{b}})}]{ginges16pra}%
  \BibitemOpen
  \bibfield  {author} {\bibinfo {author} {\bibfnamefont {J.~S.~M.}\
  \bibnamefont {Ginges}}\ and\ \bibinfo {author} {\bibfnamefont {J.~C.}\
  \bibnamefont {Berengut}},\ }\href {\doibase 10.1103/PhysRevA.93.052509}
  {\bibfield  {journal} {\bibinfo  {journal} {Phys. Rev. A}\ }\textbf {\bibinfo
  {volume} {93}},\ \bibinfo {pages} {052509} (\bibinfo {year}
  {2016}{\natexlab{b}})}\BibitemShut {NoStop}%
\bibitem [{\citenamefont {Flambaum}\ and\ \citenamefont
  {Ginges}(2005)}]{flambaum05pra}%
  \BibitemOpen
  \bibfield  {author} {\bibinfo {author} {\bibfnamefont {V.~V.}\ \bibnamefont
  {Flambaum}}\ and\ \bibinfo {author} {\bibfnamefont {J.~S.~M.}\ \bibnamefont
  {Ginges}},\ }\href {\doibase 10.1103/PhysRevA.72.052115} {\bibfield
  {journal} {\bibinfo  {journal} {Phys. Rev. A}\ }\textbf {\bibinfo {volume}
  {72}},\ \bibinfo {pages} {052115} (\bibinfo {year} {2005})}\BibitemShut
  {NoStop}%
\bibitem [{\citenamefont {Parpia}\ \emph {et~al.}(1996)\citenamefont {Parpia},
  \citenamefont {Froese~Fischer},\ and\ \citenamefont {Grant}}]{Parpia1996}%
  \BibitemOpen
  \bibfield  {author} {\bibinfo {author} {\bibfnamefont {F.~A.}\ \bibnamefont
  {Parpia}}, \bibinfo {author} {\bibfnamefont {C.}~\bibnamefont
  {Froese~Fischer}}, \ and\ \bibinfo {author} {\bibfnamefont {I.~P.}\
  \bibnamefont {Grant}},\ }\href {\doibase 10.1016/0010-4655(95)00136-0}
  {\bibfield  {journal} {\bibinfo  {journal} {Comp. Phys. Comm.}\ }\textbf
  {\bibinfo {volume} {94}},\ \bibinfo {pages} {249} (\bibinfo {year}
  {1996})}\BibitemShut {NoStop}%
\bibitem [{\citenamefont {Hansen}\ \emph
  {et~al.}(1988{\natexlab{b}})\citenamefont {Hansen}, \citenamefont {Raassen},
  \citenamefont {Uylings},\ and\ \citenamefont {Lister}}]{Hansen1988b}%
  \BibitemOpen
  \bibfield  {author} {\bibinfo {author} {\bibfnamefont {J.}~\bibnamefont
  {Hansen}}, \bibinfo {author} {\bibfnamefont {A.}~\bibnamefont {Raassen}},
  \bibinfo {author} {\bibfnamefont {P.}~\bibnamefont {Uylings}}, \ and\
  \bibinfo {author} {\bibfnamefont {G.}~\bibnamefont {Lister}},\ }\href
  {\doibase 10.1016/0168-583X(88)90405-3} {\bibfield  {journal} {\bibinfo
  {journal} {Nucl. Instrum. Meth. Phys. Res. Sect. B: Beam Interact. Mater.
  Atoms}\ }\textbf {\bibinfo {volume} {31}},\ \bibinfo {pages} {134} (\bibinfo
  {year} {1988}{\natexlab{b}})}\BibitemShut {NoStop}%
\bibitem [{\citenamefont {Ryabtsev}\ and\ \citenamefont
  {Kononov}(2012)}]{Ryabtsev2012}%
  \BibitemOpen
  \bibfield  {author} {\bibinfo {author} {\bibfnamefont {A.~N.}\ \bibnamefont
  {Ryabtsev}}\ and\ \bibinfo {author} {\bibfnamefont {E.~Y.}\ \bibnamefont
  {Kononov}},\ }\href {\doibase 10.1088/0031-8949/85/02/025301} {\bibfield
  {journal} {\bibinfo  {journal} {Phys. Scr.}\ }\textbf {\bibinfo {volume}
  {85}},\ \bibinfo {pages} {025301} (\bibinfo {year} {2012})}\BibitemShut
  {NoStop}%
\bibitem [{\citenamefont {Van~het Hof}\ and\ \citenamefont
  {Joshi}(1993)}]{VanhetHof1993}%
  \BibitemOpen
  \bibfield  {author} {\bibinfo {author} {\bibfnamefont {G.}~\bibnamefont
  {Van~het Hof}}\ and\ \bibinfo {author} {\bibfnamefont {Y.}~\bibnamefont
  {Joshi}},\ }\href {\doibase 10.1088/0031-8949/48/6/010} {\bibfield  {journal}
  {\bibinfo  {journal} {Phys. Scr.}\ }\textbf {\bibinfo {volume} {48}},\
  \bibinfo {pages} {714} (\bibinfo {year} {1993})}\BibitemShut {NoStop}%
\bibitem [{\citenamefont {Azarov}\ \emph {et~al.}(1994)\citenamefont {Azarov},
  \citenamefont {Joshi}, \citenamefont {Churilov},\ and\ \citenamefont
  {Ryabtsev}}]{azarov1994analysis}%
  \BibitemOpen
  \bibfield  {author} {\bibinfo {author} {\bibfnamefont {V.}~\bibnamefont
  {Azarov}}, \bibinfo {author} {\bibfnamefont {Y.}~\bibnamefont {Joshi}},
  \bibinfo {author} {\bibfnamefont {S.}~\bibnamefont {Churilov}}, \ and\
  \bibinfo {author} {\bibfnamefont {A.}~\bibnamefont {Ryabtsev}},\ }\href
  {\doibase 10.1088/0031-8949/50/6/008} {\bibfield  {journal} {\bibinfo
  {journal} {Phys. Scr.}\ }\textbf {\bibinfo {volume} {50}},\ \bibinfo {pages}
  {642} (\bibinfo {year} {1994})}\BibitemShut {NoStop}%
\bibitem [{\citenamefont {Kramida}(2011)}]{Kramida2011}%
  \BibitemOpen
  \bibfield  {author} {\bibinfo {author} {\bibfnamefont {A.}~\bibnamefont
  {Kramida}},\ }\href {\doibase 10.1016/j.cpc.2010.09.019} {\bibfield
  {journal} {\bibinfo  {journal} {Comp. Phys. Comm.}\ }\textbf {\bibinfo
  {volume} {182}},\ \bibinfo {pages} {419} (\bibinfo {year}
  {2011})}\BibitemShut {NoStop}%
\bibitem [{\citenamefont {{J. R. Crespo L\'{o}pez-Urrutia}}\ \emph
  {et~al.}(2002)\citenamefont {{J. R. Crespo L\'{o}pez-Urrutia}}, \citenamefont
  {Beiersdorfer}, \citenamefont {Widmann},\ and\ \citenamefont
  {Decaux}}]{lopez2002visible}%
  \BibitemOpen
  \bibfield  {author} {\bibinfo {author} {\bibnamefont {{J. R. Crespo
  L\'{o}pez-Urrutia}}}, \bibinfo {author} {\bibfnamefont {P.}~\bibnamefont
  {Beiersdorfer}}, \bibinfo {author} {\bibfnamefont {K.}~\bibnamefont
  {Widmann}}, \ and\ \bibinfo {author} {\bibfnamefont {V.}~\bibnamefont
  {Decaux}},\ }\href {\doibase 10.1139/p02-080} {\bibfield  {journal} {\bibinfo
   {journal} {Can. J. Phys.}\ }\textbf {\bibinfo {volume} {80}},\ \bibinfo
  {pages} {1687} (\bibinfo {year} {2002})}\BibitemShut {NoStop}%
\bibitem [{\citenamefont {Rodrigues}\ \emph {et~al.}(2004)\citenamefont
  {Rodrigues}, \citenamefont {Indelicato}, \citenamefont {Santos},
  \citenamefont {Patt{\'e}},\ and\ \citenamefont
  {Parente}}]{Rodrigues2004systematic}%
  \BibitemOpen
  \bibfield  {author} {\bibinfo {author} {\bibfnamefont {G.~C.}\ \bibnamefont
  {Rodrigues}}, \bibinfo {author} {\bibfnamefont {P.}~\bibnamefont
  {Indelicato}}, \bibinfo {author} {\bibfnamefont {J.~P.}\ \bibnamefont
  {Santos}}, \bibinfo {author} {\bibfnamefont {P.}~\bibnamefont {Patt{\'e}}}, \
  and\ \bibinfo {author} {\bibfnamefont {F.}~\bibnamefont {Parente}},\ }\href
  {\doibase 10.1016/j.adt.2003.11.005} {\bibfield  {journal} {\bibinfo
  {journal} {At. Data. Nucl. Data Tables}\ }\textbf {\bibinfo {volume} {86}},\
  \bibinfo {pages} {117} (\bibinfo {year} {2004})}\BibitemShut {NoStop}%
\bibitem [{\citenamefont {Kramida}\ \emph {et~al.}(2015)\citenamefont
  {Kramida}, \citenamefont {{Yu.~Ralchenko}}, \citenamefont {Reader},\ and\
  \citenamefont {{and NIST ASD Team}}}]{NIST_ASD}%
  \BibitemOpen
  \bibfield  {author} {\bibinfo {author} {\bibfnamefont {A.}~\bibnamefont
  {Kramida}}, \bibinfo {author} {\bibnamefont {{Yu.~Ralchenko}}}, \bibinfo
  {author} {\bibfnamefont {J.}~\bibnamefont {Reader}}, \ and\ \bibinfo {author}
  {\bibnamefont {{and NIST ASD Team}}},\ }\href@noop {} {}\bibinfo
  {howpublished} {{NIST Atomic Spectra Database (ver. 5.3), [Online].
  Available: {\tt{http://physics.nist.gov/asd}} [2016, February 2]. National
  Institute of Standards and Technology, Gaithersburg, MD.}} (\bibinfo {year}
  {2015})\BibitemShut {NoStop}%
\bibitem [{\citenamefont {Nielson}\ and\ \citenamefont
  {Koster}(1963)}]{nielson1963}%
  \BibitemOpen
  \bibfield  {author} {\bibinfo {author} {\bibfnamefont {C.~W.}\ \bibnamefont
  {Nielson}}\ and\ \bibinfo {author} {\bibfnamefont {G.~F.}\ \bibnamefont
  {Koster}},\ }\href@noop {} {\emph {\bibinfo {title} {Spectroscopic
  Coefficients for the p$^n$, d$\,^n$, and f$\,^n$ Configurations}}}\ (\bibinfo
   {publisher} {MIT press},\ \bibinfo {year} {1963})\BibitemShut {NoStop}%
\bibitem [{\citenamefont {Raassen}\ and\ \citenamefont {van
  Kleef}(1986)}]{Raassen1986}%
  \BibitemOpen
  \bibfield  {author} {\bibinfo {author} {\bibfnamefont {A.}~\bibnamefont
  {Raassen}}\ and\ \bibinfo {author} {\bibfnamefont {T.~A.}\ \bibnamefont {van
  Kleef}},\ }\href {\doibase 10.1016/0378-4363(86)90030-6} {\bibfield
  {journal} {\bibinfo  {journal} {Physica B+C}\ }\textbf {\bibinfo {volume}
  {142}},\ \bibinfo {pages} {359} (\bibinfo {year} {1986})}\BibitemShut
  {NoStop}%
\bibitem [{\citenamefont {Raassen}\ and\ \citenamefont
  {Van~Kleef}(1987)}]{Raassen1987}%
  \BibitemOpen
  \bibfield  {author} {\bibinfo {author} {\bibfnamefont {A.}~\bibnamefont
  {Raassen}}\ and\ \bibinfo {author} {\bibfnamefont {T.~A.}\ \bibnamefont
  {Van~Kleef}},\ }\href {\doibase 10.1016/0378-4363(87)90134-3} {\bibfield
  {journal} {\bibinfo  {journal} {Physica B+C}\ }\textbf {\bibinfo {volume}
  {146}},\ \bibinfo {pages} {423} (\bibinfo {year} {1987})}\BibitemShut
  {NoStop}%
\bibitem [{\citenamefont {Ryabtsev}\ and\ \citenamefont
  {Kononov}(2016)}]{Ryabtsev2016}%
  \BibitemOpen
  \bibfield  {author} {\bibinfo {author} {\bibfnamefont {A.~N.}\ \bibnamefont
  {Ryabtsev}}\ and\ \bibinfo {author} {\bibfnamefont {E.~Y.}\ \bibnamefont
  {Kononov}},\ }\href {\doibase 10.1088/0031-8949/91/2/025402} {\bibfield
  {journal} {\bibinfo  {journal} {Phys. Scr.}\ }\textbf {\bibinfo {volume}
  {91}},\ \bibinfo {pages} {025402} (\bibinfo {year} {2016})}\BibitemShut
  {NoStop}%
\bibitem [{\citenamefont {Ryabtsev}\ and\ \citenamefont
  {Kononov}(2011)}]{Ryabtsev2011}%
  \BibitemOpen
  \bibfield  {author} {\bibinfo {author} {\bibfnamefont {A.~N.}\ \bibnamefont
  {Ryabtsev}}\ and\ \bibinfo {author} {\bibfnamefont {E.~Y.}\ \bibnamefont
  {Kononov}},\ }\href {\doibase 10.1088/0031-8949/84/01/015301} {\bibfield
  {journal} {\bibinfo  {journal} {Phys. Scr.}\ }\textbf {\bibinfo {volume}
  {84}},\ \bibinfo {pages} {015301} (\bibinfo {year} {2011})}\BibitemShut
  {NoStop}%
\bibitem [{\citenamefont {Van~Kleef}\ \emph {et~al.}(1987)\citenamefont
  {Van~Kleef}, \citenamefont {Raassen},\ and\ \citenamefont
  {Joshi}}]{VanKleef1987}%
  \BibitemOpen
  \bibfield  {author} {\bibinfo {author} {\bibfnamefont {T.~A.}\ \bibnamefont
  {Van~Kleef}}, \bibinfo {author} {\bibfnamefont {A.}~\bibnamefont {Raassen}},
  \ and\ \bibinfo {author} {\bibfnamefont {Y.}~\bibnamefont {Joshi}},\ }\href
  {\doibase 10.1088/0031-8949/36/1/023} {\bibfield  {journal} {\bibinfo
  {journal} {Phys. Scr.}\ }\textbf {\bibinfo {volume} {36}},\ \bibinfo {pages}
  {140} (\bibinfo {year} {1987})}\BibitemShut {NoStop}%
\bibitem [{\citenamefont {Joshi}\ \emph {et~al.}(1988)\citenamefont {Joshi},
  \citenamefont {Raassen}, \citenamefont {Van~Kleef},\ and\ \citenamefont
  {Van~der Valk}}]{Joshi1988}%
  \BibitemOpen
  \bibfield  {author} {\bibinfo {author} {\bibfnamefont {Y.}~\bibnamefont
  {Joshi}}, \bibinfo {author} {\bibfnamefont {A.}~\bibnamefont {Raassen}},
  \bibinfo {author} {\bibfnamefont {T.~A.}\ \bibnamefont {Van~Kleef}}, \ and\
  \bibinfo {author} {\bibfnamefont {A.}~\bibnamefont {Van~der Valk}},\ }\href
  {\doibase 10.1088/0031-8949/38/5/007} {\bibfield  {journal} {\bibinfo
  {journal} {Phys. Scr.}\ }\textbf {\bibinfo {volume} {38}},\ \bibinfo {pages}
  {677} (\bibinfo {year} {1988})}\BibitemShut {NoStop}%
\bibitem [{\citenamefont {Ryabtsev}(2016)}]{Ryabtsev2016b}%
  \BibitemOpen
  \bibfield  {author} {\bibinfo {author} {\bibfnamefont {A.~N.}\ \bibnamefont
  {Ryabtsev}},\ }\href@noop {} {}\bibinfo {howpublished} {in preparation}
  (\bibinfo {year} {2016})\BibitemShut {NoStop}%
\bibitem [{\citenamefont {Kramida}(2010)}]{Kramida2010}%
  \BibitemOpen
  \bibfield  {author} {\bibinfo {author} {\bibfnamefont {A.~E.}\ \bibnamefont
  {Kramida}},\ }\href@noop {} {\bibfield  {journal} {\bibinfo  {journal}
  {Comput. Phys. Commun.}\ }\textbf {\bibinfo {volume} {182}},\ \bibinfo
  {pages} {419} (\bibinfo {year} {2010})}\BibitemShut {NoStop}%
\end{thebibliography}%
\end{document}